\documentclass{jfm}
\usepackage{mathrsfs}

\usepackage{latexsym,amssymb,lastpage}
\usepackage{graphicx,amsfonts}
\usepackage{times,mathptmx,bm,amsmath}
\usepackage{dcolumn}

\newcommand\Real{\mbox{Re}}

\newcommand{\df}[2]{\displaystyle\frac{#1}{#2}}
\newcommand{\tf}[2]{\textstyle\frac{#1}{#2}}
\newcommand{\Ssf}[2]{\displaystyle\sum_{#1}^{#2}}
\newcommand{\Int}[2]{\displaystyle\int_{#1}^{#2}}

\newcommand{\PDD}[2]{\df{\partial #1}{\partial #2}}
\newcommand{\PD}[1]{\df{\partial }{\partial #1}}

\newcommand{\PDDT}[2]{\df{\partial^{2} #1}{\partial #2^{2}}}

\newcommand{\PTT}[1]{\df{\partial^{2} }{\partial #1^{2}}}

\newcommand{\os}[1]{\overline{#1}}

\newcommand{\ts}[1]{\tilde{#1}}

\newcommand{\eps}{\varepsilon}

\newcommand{\be}{\begin{eqnarray}}
\newcommand{\en}{\end{eqnarray}}
\newcommand{\no}{\nonumber}

\newcommand\shalf{\ensuremath{{\scriptstyle\frac{1}{2}}}}
\newcommand\sthird{\ensuremath{{\scriptstyle\frac{1}{3}}}}

\newcommand\etal{\mbox{\textit{et al.}}}

\newcommand{\beqa}{\begin{eqnarray}}
\newcommand{\eeqa}{\end{eqnarray}}

\def\Xint#1{\mathchoice
{\XXint\displaystyle\textstyle{#1}}%
{\XXint\textstyle\scriptstyle{#1}}%
{\XXint\scriptstyle\scriptscriptstyle{#1}}%
{\XXint\scriptscriptstyle\scriptscriptstyle{#1}}%
\!\int}
\def\XXint#1#2#3{{\setbox0=\hbox{$#1{#2#3}{\int}$}
\vcenter{\hbox{$#2#3$}}\kern-.5\wd0}}

\def\dashint{\Xint-}

\begin{document}

\title[Asymptotic analysis for generation of unsteady waves by turbulence]
{An asymptotic analysis for generation of unsteady surface waves on deep water by turbulence}

\author[S.G. Sajjadi]{S.G. Sajjadi}

\affiliation{Department of Mathematics, ERAU, FL, USA.\\
and Trinity College, Cambridge, UK.}

\label{firstpage}

\maketitle

\begin{abstract}{The detailed mathematical study of the recent paper by Sajjadi, Hunt and Drullion (2014) is presented. The mathematical developement considered by them, for  unsteady growing monochromatic waves is also extended to Stokes waves. The present contribution also demonstrates agreement with the pioneering work of Belcher and Hunt (1993) which is valid in the limit of the complex part of the wave phase speed $c_i\downarrow 0$. It is further shown that the energy-transfer parameter and the surface shear stress for a Stokes wave reverts to a monochromatic wave when the second harmonic is excluded. Furthermore, the present theory can be used to estimate the amount of energy 
transferred to each component of nonlinear surface waves on deep water from a turbulent shear flow blowing over it. Finally, it is demonstrated that in the presence of turbulent eddy viscosity the Miles (1957) critical layer does not play an important role. Thus, it is concluded that in the limit of zero growth rate the effect of the wave growth arises from the elevated critical layer by finite turbulent diffusivity, so that the perturbed flow and the drag force is determined by the asymmetric and sheltering flow in the surface shear layer and its matched interaction with the upper region.}

\end{abstract}

\section{Introduction}

It is well known that a surface wave travelling along a water surface can force a couple motion in the air and water, both propagating at the same speed, namely the eigenvalue $c_r$, being the real part of the wave speed $c$. Hence the surface wave could force an unstable shear mode in the air, which then grows and induces growth of the water wave. In a pioneering work, Miles (1957) constructed a model for generation of waves by shear flows by assuming that the critical height is sufficiently high that the turbulent stresses could be neglected. Given this assumption, he argued that the airflow perturbations are described by Rayleigh equation 
\be
(U-c)(\phi''-k^2\phi)-U''\phi=0\label{jwm1}
\en
for the non-dimensional perturbation stream function $\phi$. In (\ref{jwm1}), $U(z)$ is the undisturbed velocity profile for the
wind, blowing over the waves, and $k$ is the wavenumber.
Clearly, unless the wave amplitude varies with time, i.e. $c_i\neq 0$, the equation (\ref{jwm1}) is singular at critical point $z_c$. By solving (\ref{jwm1}) above and below the air-water interface and by matching the vertical velocity and pressure at $z_c$, Miles (1957) calculated $c_i$ in the limit as $c_i/U_*\downarrow 0$ from the resulting eigenvalue relationship.

 Miles (1957) showed
\be
c^2=c_w^2+s(\alpha+i\beta)U_1^2\label{in1}
\en
where $c_w=\sqrt{g/k}$ is the free surface wave speed,
$s=\rho_a/\rho_w\ll 1$
and $U_1=U_*/\kappa$, with $U_*$ representing the wind friction velocity, $\rho_a$
and $\rho_w$ are
air and water densities, respectively, and $\kappa$ is the von K\'arm\'an's
constant.
He then deduced the growth of a monochromatic wave is given by the following
expression
\be
\zeta_a=2{\mathscr I}\{c\}/{\mathscr R}\{c\}=s\beta(U_1/c)\label{in2}
\en
The very important aspect of (\ref{in1}) and (\ref{in2}) is that, for a
steady wave
(in which the wave amplitude $a$ remains constant), $c$ must have a non-zero
imaginary part,
i.e. such a wave will only grow if $c_i\neq 0$, This is quite evident from
equations
(\ref{in1}) and (\ref{in2}).

In his paper Miles derived an integral expression for, what is commonly known as,
the `energy-transfer parameter' $\beta$,
\be
\beta=-{\mathscr I}\left\{\int_{\eta_c}^{\infty}|\phi|^2(w''/w)\,d\eta\right\}\label{in3}
\en
where $w$ is the dimensionless wind velocity profile and the suffix $c$
indicating evaluation at the critical point $\eta=\eta_c$ where the
wind velocity equals the wave speed. However, in evaluation the
integral in (\ref{in3}) and hence arriving at his `well known'
inviscid expression for $\beta$ at $\eta=\eta_c$, for $w_c=0$ and $c_i\neq 0$,
\be
\beta=-\pi|\phi_c|^2(w_c''/w_c'),\label{in4}
\en
he states ``The path of integration in (\ref{in4}) must be
indented either over or under the singularity at $\eta=\eta_c$,
where $w(\eta_c)=0$, and on this choice depends the sign of
$\beta...$''. At the bottom of the same paragraph he concludes
that ``... the path must be indented under the singularity$^*$.''
Note the asterisks attached to the word `singularity' refers to a
crucial footnote which holds the key to Miles' (1957)
critical-layer theory. In this footnote he states ``$^*$ This
assumes $c=c_w$ is real. In the next approximation ${\mathscr I}\{c\}>0$,
so that the singularity lies slightly above the real axis
(assuming $w_c''/w_c<0$), and the path of integration in
(\ref{in3}) passes under the singularity without the necessity of
indentation.''

This important footnote is generally overlooked by many who
refer to Miles' (1957) critical-layer mechanism. This footnote is
very significant in that (a) has a physical consequence which
contradicts the results in
(\ref{in1}) and (\ref{in2}) which shows clearly if $c_i=0$ then
$\beta=0$ and hence waves cannot grow; and (b) has a mathematical 
consequence which indicates that equation (\ref{in4}) is valid if 
$c_i\neq 0$.

\subsection{Physical mechanisms}

We now present two alternative physical arguments that not only their
results agree with each other but proves rigoriously that Miles (1957)
critical-layer mechanism is valid only for slowly-growing waves which
does not apply to growth of surface waves by strong or turbulent shear
flows in open ocean.

Belcher \& Hunt (1993) (referred to as BH therein) considered a fully developed
boundary layer over a two-dimensional monochromatic wave of small steepness $ak$
propagating with small wave speed $c$ and calculated the perturbations in the
asymptotic limit ${\cal U}\equiv(U_*+c)/U\downarrow 0$. In this limit, the
critical height $z_c$ lies within the inner surface layer, where the perturbation
Reynolds shear stress varies slowly. Then, by considering the equation for the
shear stress, they constructed solutions across the critical layer and demonstrated
that the shear stress perturbation plays an important role at the critical height
which in turn implies Miles' (1957) inviscid theory is {\em not} the dominating
mechanism for the wave growth in this parameter range. Note that the perturbations
above the inner surface layer are not directly influenced by the critical height and
the region below $z_c$ where the flow reversal occurs (see figure 1). In fact this is
very similar to the perturbations due to a static undulation, but with the difference
that the effective roughness length, that determines the shape of the unperturbed
velocity profile, is modified according to
$$z_c=z_0\exp(\kappa c/U_*).$$
BH then used the solutions for the perturbations to the boundary layer and calculated
the wave growth, which is determined, in the leading order of perturbation, by the 
asymmetric pressure perturbation induced by the thickening of the perturbed boundary
layer on the leeside of the wave crest. To the first order in ${\cal U}$, BH discovered
that there are new effects that contribute significantly to the rate of growth:
(a) the asymmetries in both the normal and shear Reynolds shear stresses associated
with the leeside thickening of the boundary layer, this they termed the {\em non-separated
sheltering} (cf. Jeffreys 1925); (b) asymmetrical perturbations which are induced by the
varying surface velocity associated with the fluid motion in the wave; and, (c) 
asymmetries induced by the variation in the surface roughness along the wave. The theoretical
value, predicted by their theory, for the shear stress perturbation at the crest of the wave
on the wave surface as well as on the top of the inner region is in good agreement with
laboratory measurements. Hence, despite the restriction that ${\cal U}\ll 1$, their theory
describes a large portion of the experimental observations of the wave growth rate made
at sea and in the laboratory.

\begin{figure}
   \begin{center}
\includegraphics[width=10cm]{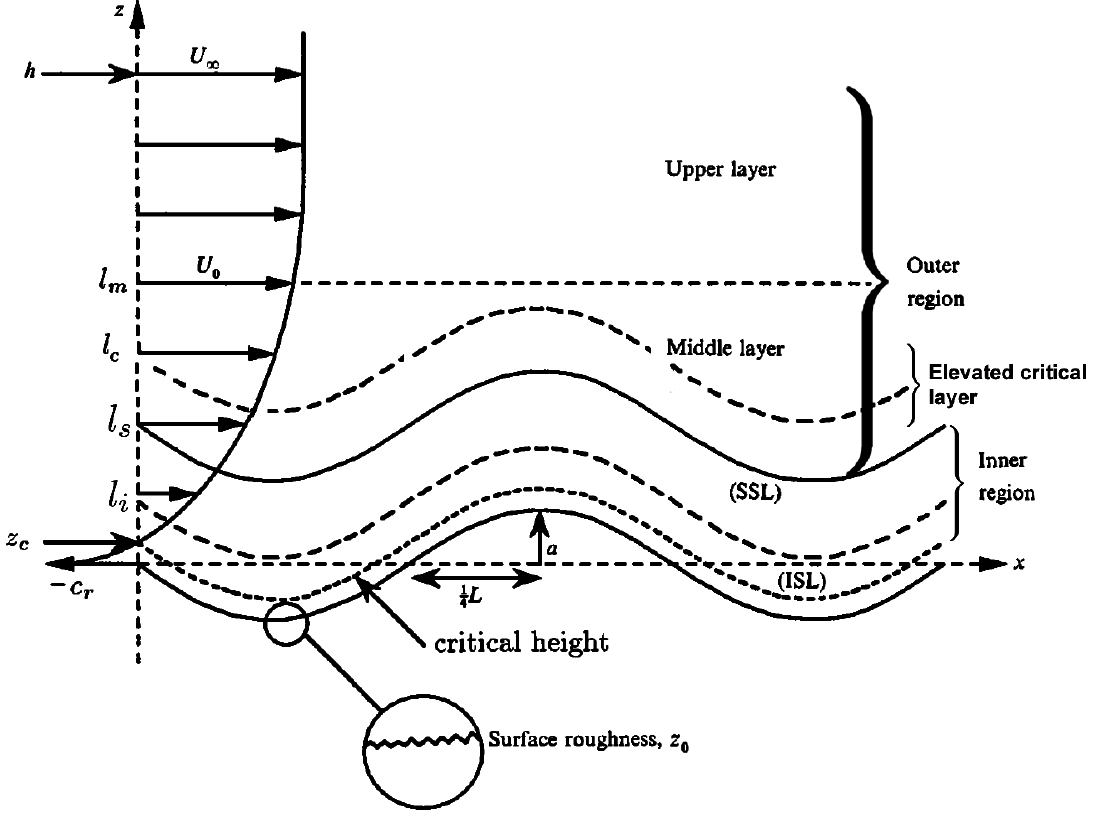}
\caption{\footnotesize Schematic diagram for flow geometry and asymptotic multi-layer structure for analysing  turbulent shear flow over steady and unsteady monochromatic waves. Taken from Sajjadi {\em et al.} (2014).}
   \end{center}
\end{figure}

In a subsequent study, Belcher \etal (1999) (hereafter will be referred to as BHC) considered 
turbulent flow over growing waves, using triple-deck boundary-layer theory originally
developed by Lighthill (1957) and Stewartson (1974), to analyse the sheltering mechanism
described above. They suggested that Miles' (1957) critical-layer theory
generates growth of waves but this was not demonstrated in data. In this model they
assumed the atmospheric mean flow is neutrally stable and is logarithmic and the surface wave,
moving in the positive $x$-direction, is monochromatic whose profile is given by
$$z'=\Real\{ae^{ik(x'-ct)}\}$$
where $c=c_r+ic_i$ is a complex wave speed, such that $c_r$ is the phase speed and the wave
amplitude, $a$, grows exponentially at the rate $kc_i$. They considered a frame of reference
moving with the wave at a speed $c_r$ so that the mean velocity profile can be expressed as
$$U(z)=U_1\log(z/z_0)-c_r$$
where $U_1=U_*/\kappa$, and that this wave speed vanishes at the critical height $z_c$. In this
frame of reference the surface wave is described by
$$z=\Real\{ae^{ik(x-ic_it)}\}.$$
The boundary conditions imposed at the wave surface is that the wind velocity is equal to the surface
velocity of the water flow, being approximated by the surface value of the orbital velocity of an
irrotational wave on deep water. The other boundary condition is that perturbations in the basic flow
vanish as $kz\uparrow\infty.$

BHC modelled the turbulent shear stress in the inner region, adjacent to the wave surface, using a
mixing length model, and in the region above this, the outer region, they invoked rapid distortion 
theory to  describe the turbulence. They showed that the depth of the inner region, $\ell_i$, may be
obtained from the following implicit relation
$$k\ell_i=\df{2\kappa^2}{|\ln(\ell_i/z_0)-\kappa c_r/U_*|}$$
where the variation of solution to this equation for $\ell_i$ as a function of $c_r/U_*$ for $kz_0=10^{-4}$
is shown in figure 2.

They further assumed the perturbations to the air flow are governed by equation
(1.7) where the turbulent stresses on the right-hand side is modelled
by an eddy viscosity. The vertical component of the velocity perturbation $\pmb{\mathscr{U}}\equiv\mathscr{U}_i=
(\mathscr{U},0,\mathscr{W})$ is expanded in the normal form
$${\mathscr W}(x,z,t)=\hat{\mathscr W}(k,z)e^{i(kx-ic_it)},$$ where the
amplitude of the perturbation, $\hat{\mathscr W}$, satisfies the inhomogeneous
Rayleigh equation
\be
\PDDT{\hat{\mathscr W}}{z}-\left(k^2+\df{U''}{U-ic_i}\right)\hat{\mathscr W}
=\df{i}{U-ic_i}\PTT{z}\left(\nu_e\PDDT{\hat{\mathscr W}}{z}\right)\label{7.1s}
\en
where $\nu_e$ is an eddy viscosity.

\begin{figure}
   \begin{center}
\includegraphics[width=12cm]{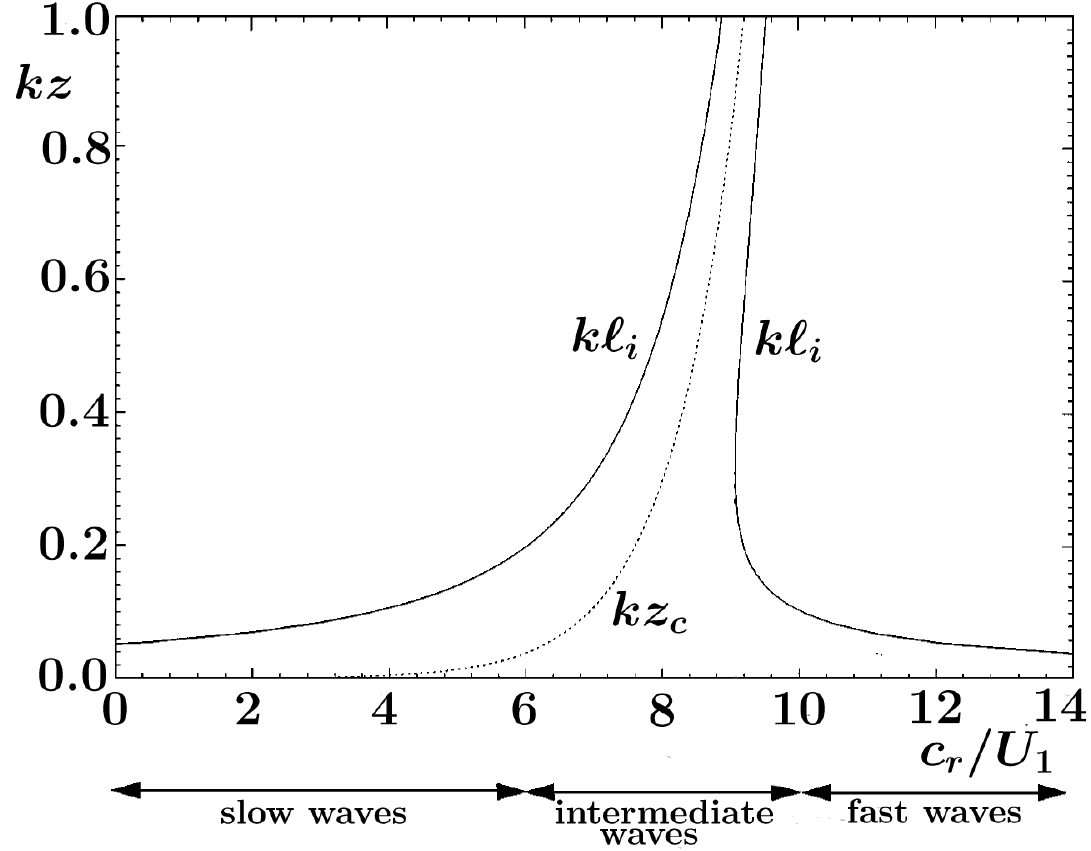}
   \end{center}
\caption{\footnotesize Variation with $c_r/U_1$ of solutions for the normalized inner-region height, $k\ell_i$, and critical
height, $kz_c$, when $kz_0 = 10^{-4}$ . For given $c_r/U1$ , an inner, local equilibrium region lies between $kz=0$ and the smallest value of $k\ell_i$, and an outer, rapid-distortion, region lies above. Solid lines: $k\ell_i$; dotted lines: $kz_c$.}
\end{figure}

BHC showed that in the middle layer, $\ell_i\ll z\ll\ell_m$,
the advection term is negligible compared with the curvature term and
the thus (\ref{7.1s}) reduces to
\be
\PDDT{\hat{\mathscr W}}{z}-\df{U''}{U-ic_i}\hat{\mathscr W}\sim 0\label{7.2s}
\en
whose leading-order solution may be expressed as
\be
\hat{\mathscr W}\sim\left\{U(z)-ic_i\right\}\left\{A+B\int\df{d\zeta}{[U(\zeta)-ic_i]^2}\right\}\label{7.3s}
\en
In (\ref{7.3s}) $A$ and $B$ are constants which may be determined by matching the inner and upper layer
solutions. Now, for slow waves, the critical layer lies close to the surface in the inner region. Hence,
the solution (\ref{7.3s}) is regular, since $U(z)>0$ and does not vanish there. Therefore, the integrand
(\ref{7.3s}) is regular throughout the middle layer. Note, the same argument applies to the moderate waves,
but there are differences in its application.

Suppose now for a range of intermediate waves the critical layer lies in the outer region. Thus, we can expect,
for a particular range of $c_r/U_*$, the critical layer lies within $\ell_i\ll z\ll\ell_m$. Neglecting the
Reynolds stresses in the vicinity of the critical layer, $\hat{\mathscr W}$ satisfies (\ref{7.2s}) with the
solution given by (\ref{7.3s}).

However, in this senario, the critical layer lies within the range of the integral in
(\ref{7.3s}) and if $c_i=0$ then there is a singularity in this integrand at the critical layer,
where $U(z_c)=0$. This singularity may be resolved by inertial effects, i.e. inviscid processes
that control its dynamics.

If we now suppose the wave grows so that $c_i>0$, then the integral (1.8) is regular at $z=z_c$.
If we further assume the waves grows slowly, such that $c_i\ll U_c^{\prime 2}/U''_c$, where the
suffix $c$ indicates evaluation at $z=z_c$, then the integral in (\ref{7.3s}) can be
evaluated approximately. To do this, $U(z)$ is Taylor expanded in the vicinity of the critical
layer, i.e. $U(z)\sim\zeta U_c'+\tf{1}{2}\zeta^2U_c''$, where $\zeta=z-z_c$ and thus the
integral becomes
\be
I&=&\int^{z-z_c}\df{d\zeta}{[U(\zeta)-ic_i]^2}\sim\int\df{d\zeta}{\left\{\zeta U_c'+\tf{1}{2}
\zeta^2U_c''-ic_i\right\}^2}\no\\
&=&\df{1}{c_iU_c'}\int\df{d\xi}{(\xi U_c'-i)^2}\left\{1+\df{\eps}{2}\df{\xi^2}{\xi-i}
\right\}^{-2}
\label{7.4s}
\en
where $\xi=\zeta U_c'/c_i$ and $\eps=c_iU_c''/U_c^{\prime 2}\ll 1$. Note that, if $U(z)$ has a
logarithmic profile then $\eps=c_i/U_*\ll 1$ which confirms that these are slowly-growing waves.
In the limit of slow-growing waves the factor in braces can be expanded for $\eps\ll 1$ to give
\be
I&\sim&\df{1}{c_iU_c'}\int\df{d\xi}{(\xi U_c'-i)^2}\left\{1-\eps\df{\xi^2}{\xi-i}+O(\eps^2)\right\}\no\\
&=&\df{1}{c_iU_c'}\left[\df{1}{\xi-i}+\eps\left\{\df{1}{2(\xi-i)^2}-\df{2i}{\xi-i}+\ln(\xi-i)\right\}
+O(\eps^2)\right]\label{7.5s}
\en
From (\ref{7.5s}) we can deduce that far from the critical height $I$ is dominated by the logarithmic term
and hence (\ref{7.5s}) reduces to
\be
I&\sim&\df{\eps}{c_iU_c'}\ln(\xi-i)\hspace*{1cm}\mbox{as $\xi\rightarrow\pm\infty$}\no\\
&=&\df{\eps}{2c_iU_c'}\ln(\xi^2+1)+i\theta\label{7.6s}
\en
where $\theta$ is given by
\be
\tan\theta=-\xi^{-1}=-c_i/U_c'(z-z_c)\label{7.7s}
\en
For a logarithmic velocity profile $\tan\theta=\eps z_c/(z-z_c)$ and hence $\theta$ varies between
\be
\theta\rightarrow 0\hspace*{0.5cm}\mbox{as $(z-z_c)/\ell_c\rightarrow\infty$}\hspace*{0.5cm}{\rm and}
\hspace*{0.5cm}\theta\rightarrow\pi\hspace*{0.5cm}\mbox{as $(z-z_c)/\ell_c\rightarrow\infty$}\label{7.8s}
\en

The imaginary part of the integral for $z\gg z_c$ is then
\be
{\mathscr I}\{I\}\sim\eps/c_iU_c'\theta=U_c''/U_c^{\prime 3}{\sf H}(z-z_c)\hspace*{0.5cm}\mbox{as $\xi\rightarrow\pm\infty$}\label{7.9s}
\en
where ${\sf H}(z-z_c)$ is the Heaviside step function. The result given by (\ref{7.9s}) is remarkable
since it is independent of $c_i$ which means even for a slowly-growing wave it leads to an out of phase
component of the motion that is independent of the growth rate.

The significance of the term $iU''_c/U_c^{\prime 3}$ in the solution for $I$ is that it yields an out of phase
contribution to the vertical velocity that ultimately leads to the same contribution to the wave growth from the
critical layer as found by Miles (1957). This result shows the solution found by Miles (1957) is valid only
when the waves grow sufficiently slowly such that
\be
c_i\ll U_c'z_c\sim U_*\label{7.10s}
\en
and hence the effects of the critical layer calculated by Miles (1957)
are valid {\em only} in the limit $c_i/U_*\downarrow 0$.

BHC further argued that in the vicinity of the critical height the turbulent shear stress perturbation,
$\Delta\tau$, can be modelled by an eddy viscosity model via
\be
\Delta\tau=\nu_e\PDD{\Delta u}{z}\label{bhc1}
\en
where $\nu_e=\alpha U_*z$ is an eddy viscosity and $\alpha=2\kappa$ is a parameter which reduces (\ref{bhc1}) to the mixing length model used in the inner region. Thus, there is a layer around the
critical height whose thickness is $\ell_r$ in which the stresses cannot be neglected. By balancing
the shear stress term in (\ref{7.1s}) with the gradient term, $\partial^2\hat{\mathscr W}/\partial z^2$,
they then estimated
$$\ell_r\sim z_c(\alpha/kz_c)^{\shalf}.$$
Hence they found that the thickness of the shear stress layer surrounding the critical height is
much smaller than the critical-layer thickness, i.e. $\ell_r\ll\ell_c$, provided
$$(\alpha/kz_c)^{\shalf}\ll c_i/U_*.$$

The main conclusions arrived by BHC, which has motivated the present investigation, is summarized below.
\begin{description}
\item{(a)} From various studies and experimental data it is well known that $c_i$ varies with
governing parameters according to
$$c_i/U_*=s(U_*/c_r)\beta$$
where $s\ll 1$ is the ratio of the air density, $\rho_a$, to that of water density, $\rho_w$, and $\beta$
is the energy-transfer parameter for the wind-wave interaction. Using this expression, the ratio of the
thickness of an inertial critical layer, $\ell_c$ to the thickness of a stress-dominated critical layer, $\ell_r$, varies according to
$$\ell_c/\ell_r\sim(kz_c/\alpha)^{\sthird}s(U_*/c_r)\beta.$$
For growing waves in the ocean $s\sim 1/800$ and $\beta\approx O(25-30)$ then for intermediate waves
when $kz_c\sim 1$ and $c_r/U_*\approx O(15-25)$, we see that $\ell_c/\ell_r\approx 1/(500\alpha^{\sthird})$. We remark that for the mixing-length model $\alpha=0.8$ and thus $\ell_c/\ell_r$
is likely to be small for a fully developed turbulent flow. Hence, one is led to conclusion that in the
intermediate regime the critical layer is dominated by effects of Reynolds stress and, contrary to Miles'
(1957) conclusion, there will not be any contribution from the critical layer to the wave growth. We
further remark that for those ocean waves that are rapidly growing in time, such that $c_i/U_*\sim 1$, the
inertial effects are dominant in the  critical layer and thus the critical-layer mechanism (c.f. Miles 1957). This suggests that such `rapidly-growing' waves might occur as a wave crest moves through a wave
group. This is currently being investigated by the present authors.
\item{(b)} For slow waves, $c_r/U_*\leq 15$, the critical height $z_c$ lies within the lower part of the
inner region and thus the reverse-flow region, situated below $z_c$ and the inner layer itself plays no
significant dynamical role. In such circumstances, asymmetry in the flow is generated by the frictional
effect of the shear stress through the inner region, resulting to lower wind speeds on the downwind side of
the wave and consequently leads to a sheltering in the lee of the wave crest. This asymmetry results to
an out-of-phase pressure perturbation which subsequently yields wave growth. We emphasize that in this 
case the air flow perturbations are similar to those over a stationary undulation, but in the range
$1\leq c_r/U_*\leq 15$ the flow is similar to that over a rough surface except now the surface roughness
is now effectively $z_c$, which increases the value of $\beta$. It is to be noted that small corrections
to the velocity of $O(akc_rk\ell_i)$ due to the orbital motions at the wave surface reduce $\beta$.

For fast waves, on the other hand, the critical layer is far above the surface, $kz_c>1$, and again there is
no significant dynamical role for the wave growth. In this scenario, the air above the wave flows largely
against the wave which induces a `negative' asymmetry from sheltering. Furthermore, orbital motions at the water surface generate additional air flow perturbations that contribute comparable `negative' asymmetries.
This 'negative' asymmetry causes an out-of-phase pressure perturbation which makes waves to decay.

Finally, between the two regions discussed above, there is also an intermediate region in which $15\leq
c_r/U_*\leq 30$ and $z_c\sim\ell$. In this region, numerical simulations show that as $c_r/U_*$ increases
from slow to the fast region, the reverse flow below the critical height becomes stronger and produces a
`negative' asymmetric displacement of streamlines upwind of the crest. However, above the critical height
the asymmetric displacement is `positive' downwind of the crest, similar to that for slow waves. Moreover,
the critical-layer mechanism also displaces streamlines downwind of the crest. Therefore, as $c_r/U_*$
increases across the intermediate region, the asymmetric component of the flow peaks to its maximum and then
decreases to zero, with the wave growth following the same trends as that of the asymmetric component of
the flow.    
\end{description}
\section{Shear stress model for unsteady wave growth}
In this section we consider the perturbation Reynolds stresses in
the flow of a turbulent wind over the surface wave
\be
z=a\cos[k(x-ct)]\equiv{h}(x,{t}),\quad(ak\ll
1)\qquad c=c_r+ic_i\label{s1.1}
\en
through an interpolation between inner, mixing-length
approximation and an outer, rapid-distortion approximation. We
will show that the wind-to-wave energy transfer predicted by this
model is substantially larger than that predicted by either Miles'
(1957) quasi-laminar model (in which the perturbation Reynolds
stresses are neglected) or Townsend's viscoelastic model (Townsend
1972) and is in very good agreement with the model proposed by Belcher \& Hunt (1993).
Here we will point out, Townsend's inner approximation
differs from the conventional mixing-length approximation and
yields a ratio of perturbation shear stress to perturbation shear
that is negative for his choice of parameters.

The equations of motion for a viscous incompressible fluid may be cast in
Cartesian tensor form as
\be
\PDD{u_i}{t}+u_j\PDD{u_i}{x_j}=\df{1}{\rho}\PDD{p_{ij}}{x_j}\label{n1}
\en
\be
\PDD{u_i}{x_i}=0\label{n2}
\en
where $x_i$ denotes the Cartesian coordinate, $u_i$ a velocity component,
$p_{ij}$ component of a stress tensor and $\rho$ the fluid density.
Decomposing the variables according to
\be
u_i=U_i+u_i'+u_i'',\hspace*{1cm}p_{ij}=P_{ij}+p_{ij}'+p_{ij}''\label{n3}
\en
where $U_i+u_i'$ and $P_{ij}+p_{ij}'$ represent a solution to (\ref{n1}),
being functions of coordinates $x_1$ and $x_3$ and having mean values with
respect to $x_2$. Note that, $U_i$ and $P_{ij}$ represent the mean
components, and $u_i'$ and $p_{ij}'$ the turbulent fluctuations. In (\ref{n3})
$u_i''$ and $p_{ij}''$ represent a small perturbation with respect to the
solution of (\ref{n1})--(\ref{n2}).

Substituting (\ref{n3}) in (\ref{n1}), neglecting second order
terms in the perturbation flow, and noting the fact that the
unperturbed flow satisfies (\ref{n1}), we obtain \be
\PDD{u_i''}{t}+(U_j+u_j')\PDD{u_i''}{x_j}+u_j''\PDD{(U_i+u_i')}{x_j}=\df{1}{\rho}\PDD{p_{ij}''}{x_j}\label{n4}
\en \be \PDD{u_i''}{x_i}=0\label{n5} \en Taking the mean values
with respect to $x_3$, the results can be expressed in the
following form
\be
\PDD{\os{u_i''}}{t}+U_j\PDD{\os{u_i''}}{x_j}+\os{u_j''}\PDD{U_i}{x_j}=\df{1}{\rho}\PD{x_j}
(\os{p_{ij}''}-\os{r_{ij}''}),\label{n6}
\en
\be
\PDD{u_i''}{x_i}=0\label{n7}
\en
Invoking the equations of continuity for both $u_i'$ and $u_i''$ the perturbation Reynolds
stress may be written as
\be
\os{r_{ij}''}&=&\rho(\os{u_i'u_j''}+\os{u_j'u_i''})\label{n8}\\
&=&\rho[\os{u_i'(u_j''-\os{u_j''})}+\os{u_j'(u_i''-\os{u_i''})}]\label{n9}
\en
with (\ref{n9}) follows from (\ref{n8}) by virtue of $\os{u_i'}=0$.

If we now set $x_1=x$, $x_3=z$, $U_1=U(z)$, $U_2=U_3=0$, $\os{u_1''}=u$,
$\os{u_3''}=w$, $\os{p_{ij}''}=-\delta_{ij}\wp$, and $\os{r_{ij}''}=\rho\os{u_i'u_j'}$,
after taking the time average, we obtain the linearized,
Reynolds-averaged equations governing $u$ and $w$, the $x$
(horizontal) and ${z}$ (vertical) components of the mean
perturbation velocity, and the kinematic perturbation pressure
${\sf p}$ as
\be
u_x+w_{z}=0,\label{s2.2}
\en
\be
(U-c)u_x+U'w=-\wp_x+\sigma_x+\tau_{z},\label{s2.3a}
\en
\be
(U-c)w_x=\wp_{z}+\tau_x,\label{s2.3b}
\en
where the subscript $x$ and ${z}$ signify partial differentiation,
$U'\equiv dU/d{z}$,
\be
\wp\equiv{\sf p}+\os{w^{\prime 2}}-(\os{w^{\prime
2}})_0,\hspace*{0.35cm}\sigma\equiv-(\os{u^{\prime
2}}-\os{w^{\prime
2}})-\sigma_0,\hspace*{0.35cm}\tau\equiv-\os{u'w'}-\tau_0\label{s2.4}
\en
$(\os{w^{\prime 2}})_0, \sigma_0$ and $\tau_0$ are the unperturbed
values of $\os{w^{\prime 2}}, -(\os{u^{\prime 2}}-\os{w^{\prime
2}})$ and $-\os{u'w'}$, and $\sigma$ is Townsend's $\tau_n$.

In this paper we consider a turbulent shear flow blowing over the surface
wave (\ref{s1.1}) whose mean velocity profile is given by
\be
U({z})=(\tau_0^{1/2}/\kappa)\log({z}/z_0)\label{s2.1}
\en
where $\tau_0$ is the kinematic shear stress in the basic flow,
$\kappa$ is K\'arm\'an's constant, ${z}$ is the elevation, $c$ is
the complex phase speed of the surface wave (\ref{s1.1}).
\subsection{Energy-transport equation}

As a first step toward a Reynolds-stress closure, Townsend points out
the transport equation for the turbulent kinetic energy
$\tf{1}{2}\os{\sf{q}^2}$ may be expressed in the form
\be
(U-c)\partial_x\left(\tf{1}{2}\os{\sf{q}^2}\right)=D+G-\eps',\label{s3.1}
\en
where
\be
D=\varrho\kappa\tau^{1/2}_0\partial_{z}\left[{z}\partial_{
z}\left(\tf{1}{2}\os{{\sf q}^2}\right)\right]\label{s3.2}
\en
represents diffusion\footnote{Townsend chooses $\varrho=0.3$ but
states that `the value ... is not critical'.}
\be
G&=&-\os{u^{\prime 2}}u_x-\os{w^{\prime 2}}w_{ z}-\os{u'w'}(U'+u_{
z}+w_x)-\tau U'\label{s3.3a}\\ & =&\sigma_0u_x+\tau_0(u_{
z}+w_x)+U'\tau\label{s3.3b}
\en
represents generation (Launder \etal 1975), (\ref{s3.3b}) follows
(\ref{s3.3a}) through (\ref{s2.2}), (\ref{s2.4}) and
linearization, and $\eps'$ repesents dissipation (see below).
Townsend neglects $w_x$ in $G$, although this appears to be
inconsistent with his subsequent rapid-distortion approximation
(see below).

Townsend's approximation to the dissipation rate $\eps'$ involves
${h}$, the surface displacement (\ref{s1.1}), but this may be
eliminated through a transformation to wave-following coordinates,
in which (Miles 1996, equation (3.4))
\be
\eps'=\tf{3}{2}\tau_0U'(e/e_0),\hspace*{0.5cm}e\equiv\os{{\sf
q}^2}-e_0,\label{s3.4}
\en
and $e_0\equiv\os{{\sf q^2}}$ in the basic flow. Substituting
(\ref{s3.3b}) and (\ref{s3.4}) into (\ref{s3.1}), neglecting
diffusion (see Miles 1996, \S3), and multiplying the result by
$2a_1$, we obtain
\be
({\mathscr D}+\lambda)a_1e-2\lambda(\tau-a_1e)=2\tau_0[a_1(u_{
z}+w_x)+a_nu_x]\equiv{\mathscr A}_i\label{s3.5}
\en
where
\be
{\mathscr D}\equiv(U-c)\partial_x,\hspace*{0.25cm}\lambda\equiv
a_1U',\hspace*{0.25cm}a_1\equiv\tau_0/e_0,\hspace*{0.25cm}a_n=\sigma_0/e_0\label{s3.6}
\en
The relaxation rate $\lambda$ is a reciprocal measure of eddy life
and dominates (is dominated by) ${\mathscr D}$ in the inner (outer)
domain.
\subsection{Rapid-distortion approximation}
To determine the outer departure of $\tau/e$ and $\sigma/e$ from
their equilibrium values $a_1$ and $a_n$, Townsend posits the
rapid-distortion approximations
\be
{\mathscr D}\left\{\df{\tau_0+\tau}{e_0+e}\right\}\simeq{\mathscr
D}\left\{\df{\tau-a_1e}{e_0}\right\}\sim A_1u_{
z}+A_2w_x+A_3u_x\label{s4.1a}
\en
and
\be
{\mathscr
D}\left\{\df{\sigma_0+\sigma}{e_0+e}\right\}\simeq{\mathscr
D}\left\{\df{\sigma-a_ne}{e_0}\right\}\sim B_1u_{
z}+B_2w_x+B_3u_x\label{s4.1b}
\en
where $A_{1,2,3}$ and $B_{1,2,3}$ are `the incremental rates of
change for suddenly imposed additional distortions'.  He then `interpolates' between
(\ref{s4.1a})--(\ref{s4.1b}) and the inner domain, in which (by
hypothesis) $\tau\rightarrow a_1e$ and $\sigma\rightarrow a_ne$,
by replacing ${\mathscr D}$ by ${\mathscr D}+\lambda$:
\be
({\mathscr D}+\lambda)(\tau-a_1e)=e_0(A_1u_{
z}+A_2w_x+A_3u_x)\equiv{\mathscr A}_0\label{s4.2a}
\en
and
\be
({\mathscr D}+\lambda)(\sigma-a_ne)=e_0(B_1u_{
z}+B_2w_x+B_3u_x\equiv){\mathscr B}_0\label{s4.2b}
\en
However, the elimination of $a_1e$ between (\ref{s3.5}) and
(\ref{s4.2a}) in the inner domain (in which $|{\mathscr
D}|\ll|\lambda|$) yields
\be
\tau\rightarrow\lambda^{-1}({\mathscr A}_i+3{\mathscr A}_0),\label{s4.3}
\en
which differs from the mixing-length approximation obtained by
invoking $|{\mathscr D}|\ll\lambda$ in (\ref{s3.5}):
\be
\tau\rightarrow a_1e\rightarrow\lambda^{-1}{\mathscr
A}_i=2\nu_0(u_{z}+w_x+{\mathscr H}u_x),\label{s4.4}
\en
where
\be
{\mathscr H}\equiv
a_n/a_1,\hspace*{0.5cm}\nu_0\equiv\tau_0/U'=\kappa\tau_0^{1/2}{
z}.\label{s4.5a,b}
\en
Indeed, if (as typically assumed in the mixing-length
approximation) $|u_x, w_x|\ll|u_{z}|$, (\ref{s4.3}) implies
\be
\df{\tau}{2\nu_0u_{z}}\rightarrow 1+\df{3A_1}{2a_1^2},\label{s4.6}
\en
which reduces to the conventional mixing-length approximation for
$A_1=0$ but is negative for Townsend's values of $a_1$ and $A_1$,
$\tf{1}{6}$ and $-0.03$.
\section{Solution of boundary-value problem}
Miles-Sajjadi theory of wave generation by turbulent wind (Miles
1996, Sajjadi 1998) reduces to the solution of the
Orr-Sommerfeld-like equation
\be
(\nu_e\Phi'')''=ik[(U-c)(\Phi''-k^2\Phi)-U''\Phi]\equiv{\mathscr
T}''\label{m1}
\en
subject to the boundary conditions
\be
\Phi=ac,\hspace*{0.5cm}\Phi'=a(kc-U')\hspace*{0.5cm}\mbox{on
$\eta=0$}\label{m2}
\en
\be
\Phi,\hspace*{0.5cm}\nu_e\Phi\rightarrow 0\hspace*{0.5cm}\mbox{as
$k\eta\rightarrow\infty$}\label{m3}
\en
where $\Phi=\Phi(\eta)$, $U=U(\eta)$ and $(\,\,\,)'\equiv
d/d\eta$. In equ (\ref{m1}) $\nu_e$ is a complex eddy viscosity
given by
\be
\nu_e=2U_*^2(U'+ikV)^{-1},\hspace*{0.5cm}V\equiv(U-c)/a_1\label{m4}
\en
where $a_1$ ($\simeq\kappa^2; \kappa=0.4$ is K\'arm\'an's constant) is
Townsend's boundary-layer constant, and $a, c$ and $k$ are the
amplitude, speed and wave number of the surface wave. The velocity
profile has the logarithmic asymptote
\be
U\sim U_1\ln(\eta/z_0),\hspace*{0.5cm}(\eta\gg z_0)\label{m5}
\en
where
\be
U_1=U_*/\kappa,\hspace*{0.5cm}z_0=\Omega U_1^2/g\label{m6}
\en
$U_*$ is the wind friction velocity and $\Omega$ is Charnock's
constant. We seek the impedance
\be
\alpha+i\beta=({\mathscr P}_0+i{\mathscr T}_0)/(kaU_1^2)\label{m7}
\en
where
\be
{\mathscr P}_0=kac^2+(ik)^{-1}(\nu_e\Phi'')'\hspace*{0.25cm}{\rm and}
\hspace*{0.25cm}{\mathscr T}_0=\nu_e\Phi'',\hspace*{0.25cm}\mbox{on
$\eta=0$}\label{m8}
\en
are complex amplitudes of the wind-induced perturbation pressure
and shear stress action on the wave.
\subsection{Reduction to second-order differential equations}
It is convenient to reduce the fourth-order differential equation
(\ref{m1}) to the pair of second-order equations
\be
& &{\mathscr T}''=(ik/\nu_e)(U-c){\mathscr
T}-ik[U''+k^2(U-c)]\Phi\label{m9a}\\ & &\Phi''={\mathscr
T}/\nu_e\label{m9b}
\en
for which the respective boundary conditions are
\be
{\mathscr T}={\mathscr T}_0,\hspace*{0.5cm}{\mathscr
T}'=ik({\mathscr P}_0-kac^2)\hspace*{0.25cm}\mbox{on
$\eta=0$}\label{m10}
\en
and (\ref{m2}). Note that, ${\mathscr T}_0$ and ${\mathscr P}_0$ are
implicitly determined by the null conditions (\ref{m3}).

We render the formulation dimensionless by referring to $\eta$ to
$k^{-1}$, $c, U$ and $V$ to $U_1$, $\Phi$ to $aU_1$, ${\mathscr P}$
and ${\mathscr T}$ to $kaU_1^2$, thereby reducing (\ref{m4}),
(\ref{m9a})--(\ref{m10}) and (\ref{m2}) to
\be
& &{\mathscr T}''=\tf{1}{2}iV(U'+iV){\mathscr
T}-i(U''+U-c)\Phi\label{m11a}\\ &
&\Phi''=(2\kappa^2)^{-1}(U'+iV){\mathscr T}\label{m11b}
\en
\be
{\mathscr T}={\mathscr T}_0,\hspace*{0.25cm}{\mathscr
T}'=i({\mathscr
P}_0-c^2),\hspace*{0.25cm}\Phi=c,\hspace*{0.25cm}\Phi'=c-U',
\hspace*{0.25cm}\mbox{on $\eta=0$}.\label{m12ad}
\en
\subsection{Inner expansion for surface layer}
A constant-stress interpolation between the logarithmic profile
(\ref{m5}) and a viscous sublayer of vanishing thickness is given
by (Rotta 1950)
\be
\df{U}{U_1}=\log(\zeta+\sqrt{\zeta^2+1})-\df{\zeta}{1+\sqrt{\zeta^2+1}}\equiv\hat{U}(\zeta),\hspace*{0.5cm}
\zeta=\df{1}{2}e\df{\eta}{z_0}\label{m13ab}
\en
in which
\be
\df{dU}{d\zeta}=\df{U_1}{1+\sqrt{1+\zeta^2}}\no
\en
Note that $U''={\cal O}(U_1/z_0^2)$ for $\zeta={\cal O}(1)$ and
therefore, in contrast to the conventional Orr-Sommerfeld problem,
is not negligible near the boundary, although it does vanish at
$\zeta=0$.

The inner and outer length scales are $z_0$ (or, more
conveniently, $\hat{z}_0\equiv 2z_0/e$) and $k^{-1}$, and the
introduction of the inner variable $\zeta$ and the small
parameter
\be
\eps\equiv k\hat{z}_0=(2\Omega/e)c^{-2}\ll 1\label{m14}
\en
in (\ref{m11a}--\ref{m11b}) and (\ref{m12ad}) leads to the inner
expansions
\be
\Phi=c+\left(\eps c-\tf{1}{2}\right)\zeta+(\eps/2\kappa^2)
\left[{\mathscr T}_0\int_0^\zeta U\,d\zeta+i{\mathscr
G}(\zeta)\right]+{\cal O}(\eps^2),\label{m15a}
\en
where
\be
{\mathscr G}(\zeta)=\int_0^\zeta
d\zeta_1\int_0^{\zeta_1}U'(\zeta_2)\,d\zeta_2\int_0^{\zeta_2}\left[\tf{1}{2}(c-U+\zeta
U')-cU'\right]\,d\zeta_3,\label{m15b}
\en
and
\be
{\mathscr T}={\mathscr T}_0+i\left[\tf{1}{2}(c+U)\zeta-cU-\Int{0}{\zeta}
U\,d\zeta\right]+{\cal O}(\eps)\label{m15c}
\en
Letting $\zeta\rightarrow\infty$ in (\ref{m15a}) and (\ref{m15b})
and invoking $\zeta=\eta/\eps$, we obtain
\be
\Phi\sim
c+\left(c-\tf{1}{2}\eps^{-1}\right)\eta+\tf{1}{2}\kappa^{-1}\left[{\mathscr
T}_0\eta(U-1)+\tf{1}{4}i\eps^{-1}\eta^2\left(c-U+\tf{7}{2}\right)\right].\label{m16}
\en
Hence, the problem is reduced in determining ${\mathscr T}$ and ${\mathscr
T}_0$. The form given by (\ref{m15c}) is not very convenient for
this task. A better approach is to determine ${\mathscr T}$ in the
inner via shear-stress-layer approximation.

In this region, the complex amplitude of the wind-induced
perturbation shear stress may be expressed as (Sajjadi 1998)
\be
{\mathscr T}=\nu_e[{\mathscr U}(\Phi/{\mathscr
U})''-2U'(\Phi/{\mathscr U})'],\hspace*{0.5cm}{\mathscr U}\equiv
U-c\label{a}
\en
Rearranging this equation, we obtain
\be
\df{{\mathscr U}{\mathscr T}}{\nu_e}={\mathscr U}\Phi''-U''\Phi.\label{b}
\en
Under the assumption that $k\eta\ll 1$ in this region, we may
neglect $k^2{\mathscr U}\Phi$ in (\ref{m1}) and upon combining the
result with (\ref{b}) we arrive at the shear-stress-layer
approximation
\be
{\mathscr T}''-(ik{\mathscr U}/\nu_e){\mathscr T}=0;\hspace*{0.5cm}(k\eta\ll
1) \label{saj}
\en
whose asymptotic solution for $k\eta\downarrow 0$, which may be expressed in terms of
of modified Bessel function of the first kind (Sajjadi 2010), is
$$ \mathscr{T}(\eta)=-2\left\{1+4\eta{\rm K}'_0(\eta)\right\}+O (\eps),$$
(cf. Appendix A of BH)\footnote{The asymptotic solution given here reduces to that 
given by BH in the limit as $c_i\downarrow 0$.}, whence for growing waves we obtain
$$\tau=-2e^{kc_it}\mathscr{T}(\eta)e^{ik(x-c_rt)}.$$ The real part of the complex amplitude of shear stress as a function of non-dimensional height is shown in figure 3. This figure clear that the shear stress becomes negative after decending from a maxima and then rising again. The latter part is not depicted in this figure

\begin{figure}
   \begin{center}
\includegraphics[width=12cm]{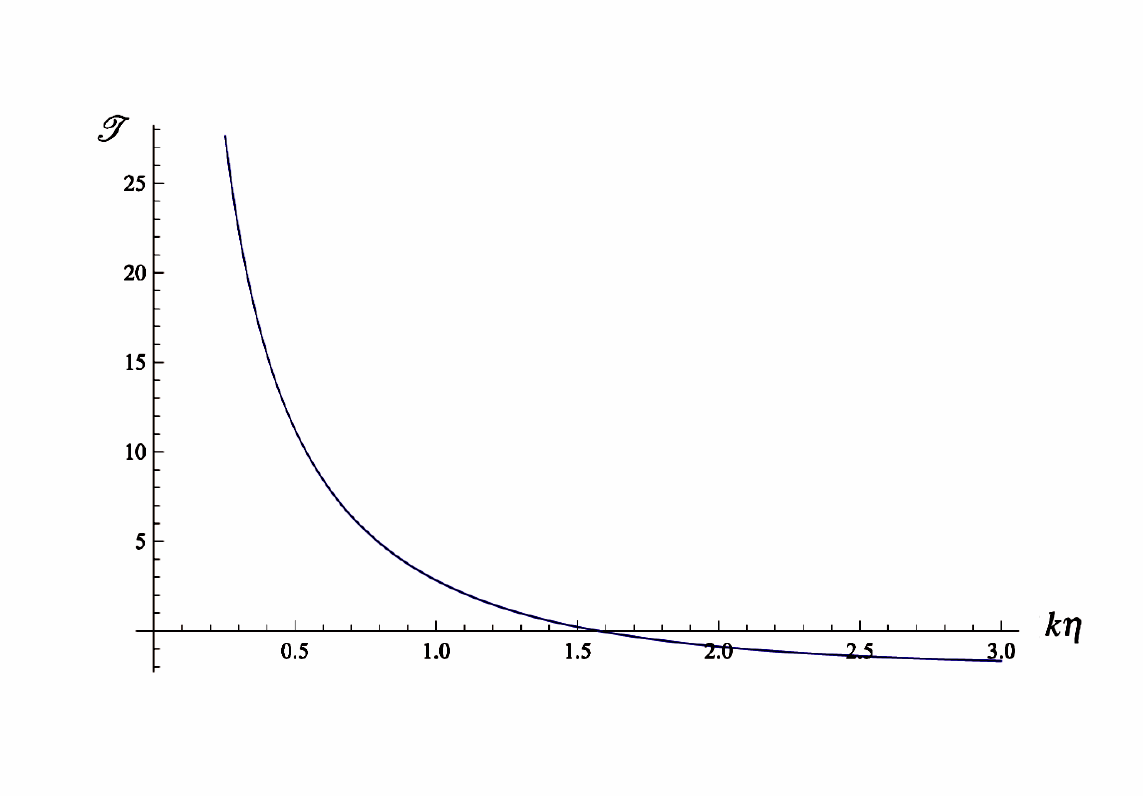}
   \end{center}
\caption{\footnotesize The real part of the complex amplitude of $\mathscr{T}$ as a function of non-dimensional height $k\eta$.}
\end{figure}

However, for the purpose of calculating the energy-transfer parameter, from
wind to wave $\beta$, we multiply (\ref{saj}) by ${\mathscr T}$ and integrate
by parts over $0<\eta<\infty$ to obtain
\be
({\mathscr T}{\mathscr T}')_0=-\int_0^\infty\left[{\mathscr T}^{\prime 2}+
(ik\mathscr{U}/\nu_e){\mathscr T}^2\right]\,d\eta\label{7.10}
\en
The integral (\ref{7.10}) is stationary with respect to first-order variations in
${\mathscr T}$ about the true solution (\ref{saj}).

Substituting the trial function
\be
{\mathscr T}={\mathscr T}_0e^{-k\eta/\delta}\label{7.11}
\en
where $\delta$ is a (complex) free parameter to be determined, and $\nu_e$
given by (\ref{m4}), with $a_1=\kappa^2$, into (\ref{7.10}) and invoking the
condition $\partial({\mathscr T}'_0/{\mathscr T_0})/\partial\delta=0$ we obtain
\be
\df{{\mathscr T}'_0}{{\mathscr T}_0}=-\df{ik}{4\kappa^2}\left[(1+\delta_*)\left(L_\delta^2
+\tf{1}{6}\pi^2\right)+2\delta^{-1}_*-\hat{c}^2\right]\label{7.12}
\en
$$
\left(L_\delta^2+2L_\delta+\tf{1}{6}\pi^2\right)\delta_*^2+2L_\delta\delta_*-2=0,\hspace*{0.5cm}
L_\delta\equiv L_0+\ln\delta,\hspace*{0.5cm}\delta_*\equiv i\kappa^{-2}\delta\eqno(3.27 {\rm a-c})
$$
where $L_0\equiv\Lambda^{-1}=-\gamma-\ln 2{\sf k}$, ${\sf k}=\Omega\hat{c}^{-2}e^{\hat{c}}$,
$\hat{c}=c/U_1$, $\gamma=0.5772$ is the Euler's number and $\Omega$ is the Charnock's constant.

Solving (3.27a) as a quadratic in $\delta_*L_\delta$ and letting $\delta_*\rightarrow 0$ we get
$$
\df{i\delta L_\delta}{\kappa^2}=\sqrt{3}-1+{\cal O}(\delta)\eqno(3.28)
$$
\section{Outer approximation above inertial critical-layer}
In contrast to the inner region, the solution in the outer region
is very straightforward. In the outer domain $k\eta\gg 1$, $U$ may
be approximated by (\ref{m5}), and $\Phi$ admits the
Green-Liouville approximation (Olver 1974, chap. 6)
\be
\Phi\sim{\mathscr F}(\mathscr Z)\exp\left[\Int{0}{\eta}{\mathscr
W}(\mathscr Z)\,d\eta\right],\hspace*{0.5cm}{\mathscr
Z}(\eta)\sim(1/\kappa^2)\ln(\eta/\eta_c)\label{m17ab}
\en
Substituting (\ref{m17ab}) into the dimensionless counterpart of
(\ref{m1}), we obtain
\be
{\mathscr W}^4+\tf{1}{2}{\mathscr Z}^2({\mathscr
W}^2-1)=0\label{m18a}
\en
and
\be
\df{{\mathscr W}({\mathscr W}^2+{\mathscr Z}^2)}{{\mathscr
F}}\df{d{\mathscr F}}{d{\mathscr Z}}+\left(6{\mathscr
W}^2+\tf{1}{2}{\mathscr Z}^2\right)\df{d{\mathscr W }}{d{\mathscr
Z}}+\df{(i\kappa^2{\mathscr W}-2){\mathscr W^3}}{\mathscr
Z}=0\label{m18b}
\en
from which it follows that
\be
{\mathscr W}^2=\tf{1}{4}\left[-{\mathscr Z}^2\pm({\mathscr
Z}^4+8{\mathscr Z}^2)^{1/2}\right],\label{m19}
\en
\be
\df{d{\mathscr W}}{d{\mathscr Z}}=\df{-{\mathscr Z}({\mathscr
W}^2-1)}{{\mathscr W}(4{\mathscr W}^2+{\mathscr
Z}^2)}=\df{2{\mathscr W}^3}{{\mathscr Z}(4{\mathscr W}^2+{\mathscr
Z}^2)}=\df{\pm 2{\mathscr W}^3}{{\mathscr Z}({\mathscr
Z}^4+8{\mathscr Z}^2)^{1/2}}\label{m20}
\en
and
\be
\df{d\ln{\mathscr F}}{d{\mathscr Z}}=-\df{{\mathscr
W}^2(4{\mathscr W}^2-{\mathscr Z}^2)}{{\mathscr Z}(4{\mathscr
W}^2+{\mathscr Z}^2)^2}-\df{i\kappa^2{\mathscr W}^3}{{\mathscr
Z}(4{\mathscr W}^2+{\mathscr Z}^2)}.\label{m21}
\en

We restrict further consideration to the asymptotic limit
$V\rightarrow\infty$, for which the admissible roots of
(\ref{m19}) may be approximated by
\be
{\mathscr W}\sim -1,\hspace*{0.5cm}-i{\mathscr
Z}/\sqrt{2}\label{m22ab}
\en
(the roots $+1$ and $+i{\mathscr Z}/\sqrt{2}$ are ruled out by the
null condition at ${\mathscr Z}=\infty$). The corresponding
approximations to (\ref{m21}) are
\be
\df{d\ln{\mathscr F}}{d{\mathscr Z}}\sim\df{1+i\kappa^2}{{\mathscr
Z}^3},\hspace*{0.5cm}-\tf{3}{2}{\mathscr
Z}^{-1}-2^{-3/2}\kappa^2,\label{m23ab}
\en
the integration of which leads, through (\ref{m17ab}), to
\be
\Phi\sim C_1e^{-\eta}+C_2{\mathscr
Z}^{-3/2}\exp\left[-\df{i\eta({\mathscr
Z}-\kappa^{-2})}{\sqrt{2}}-\df{\kappa^2{\mathscr
Z}}{2\sqrt{2}}\right], \hspace*{0.25cm}\mbox{for ${\mathscr
Z}\rightarrow\infty$}\label{m24}
\en
where $C_1$ and $C_2$ are constants. [the term
$-\tf{1}{2}{\mathscr Z}^{-2}$, derived from the real part of
(\ref{m23ab}), has been neglected in the exponent in (\ref{m24})
since it is dominated by the error implicit in the approximation
(\ref{m22ab}).] The corresponding approximation to ${\mathscr T}$,
obtained through substitution of (\ref{m24}) into (\ref{m11b}) is
\be
{\mathscr T}\sim 2i\kappa^2\left[C_1{\mathscr
Z}^{-1}e^{-\eta}-\tf{1}{2}C_2{\mathscr
Z}^{-1/2}\exp\left\{-\df{i\eta({\mathscr
Z}-\kappa^{-2})}{\sqrt{2}} -\df{\kappa^2{\mathscr
Z}}{2\sqrt{2}}\right\}\right].\label{m25}
\en

Note that, using the standard exponential substitution of the Liouville-Green
method for the asymptotic solution of (\ref{m1}), we obtain the
following expansion of the phase function
\be
\Phi&\sim &\hat{\eta}^{\gamma_1}[\log(\hat{\eta}/z_0)]^{\gamma_2}\exp[\gamma_3/\log(\hat{\eta}/z_0)]...
\exp\{i[\vartheta_1\hat{\eta}\log(\eta/z_0)\no\\
&+&\vartheta_2\hat{\eta}+\vartheta_3{\rm Ei}(1,-\log(\hat{\eta}/z_0))+\vartheta_4\hat{\eta}/\log(\hat{\eta}/z_0)+...]\}\label{8.11}
\en 
where $\vartheta_k$ and $\gamma_k$ are real,
$\hat{\eta}\approx k\eta$ and Ei is the exponential integral. Formally there is an infinite number of
terms with coefficients $\vartheta_k$ that precede the
determination of the $\gamma_k$. However, as only the first of
these, namely $\vartheta_1$, enters into the formula for
$\gamma_1$, the first two into $\gamma_2$ (although the
$\vartheta$ contributions happen to cancel), the first four into
$\gamma_3$, and so on.  The result is depicted in figure 4 where
we see that an initial expoential decay follows by an algebraic tail.
It should be noted that this is agreement with numerical simulation 
of Ierley and Miles (2001), and the expression (\ref{8.11}) is originally
found by them. However, the results of figure 4 is drawn from the
analytical expression given by the equation (\ref{8.11}).
\begin{figure}
   \begin{center}
\includegraphics[width=12cm]{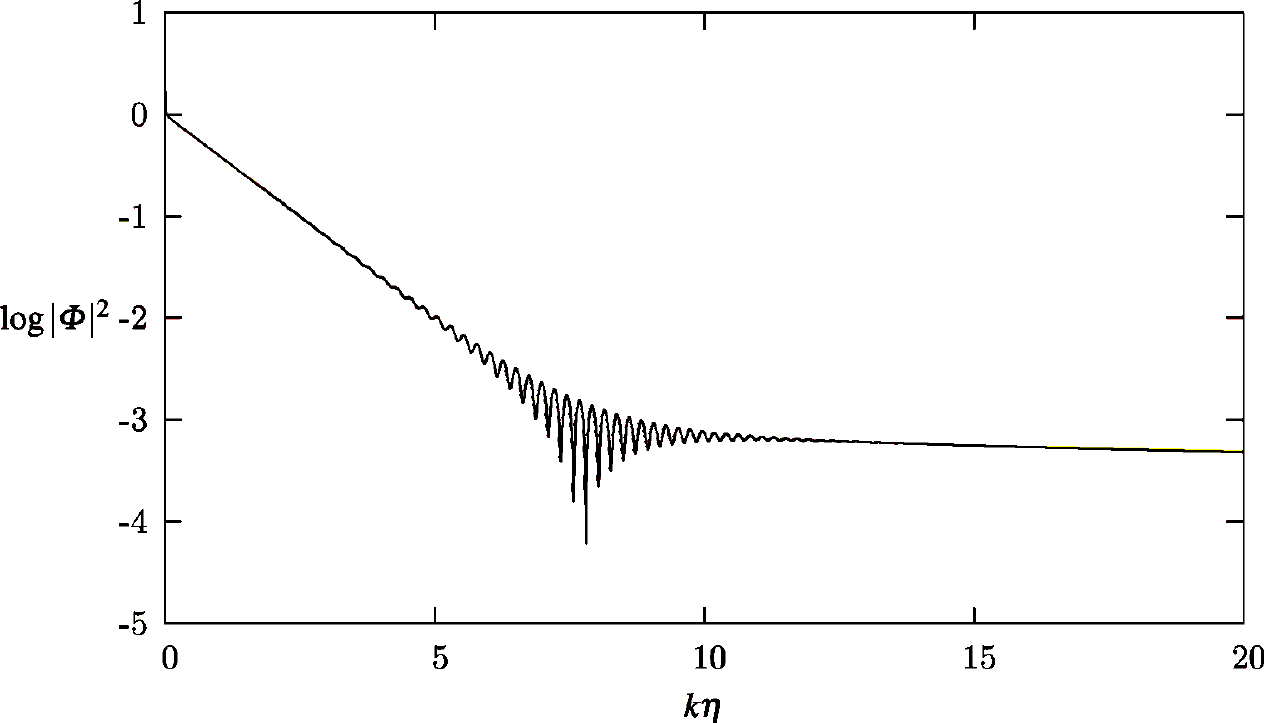}\\
\includegraphics[width=12cm]{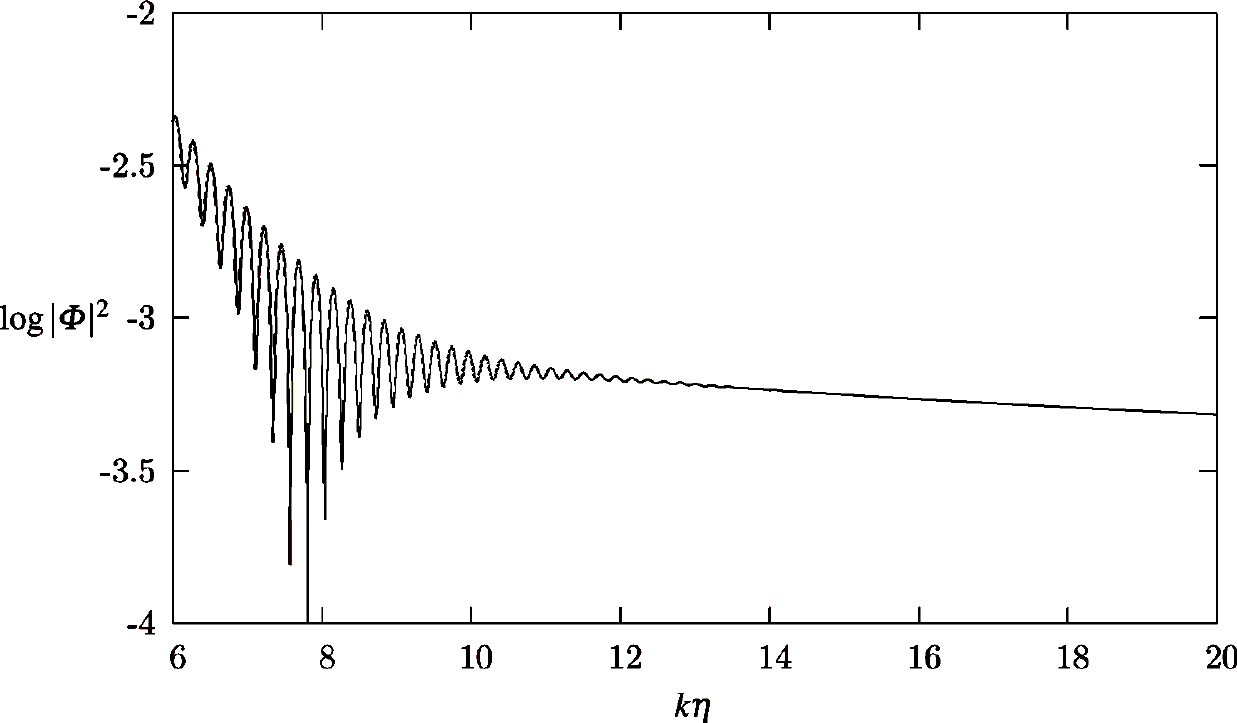}
   \end{center}
\caption{\footnotesize  The variation of the log of moduls square for perturbation stream function $\Phi$ against the non-dimentional distance $k\eta$ showing the initial exponential decay follows by an algebraic tail.
The lower figure is a close of the upper figure in the range $6\leq k\eta\leq 20$.}
\end{figure}

Since the governing equation is fourth order, we find $\vartheta$
by solving for the roots of the fourth degree polynomial
\be
\vartheta^2_1\left[(2\kappa^4+A_1)\vartheta_1^2+1\right]=0\nonumber
\en

Two roots are zero and the other two constitute an imaginary pair.
The double zero leads to a particularly simple result at next
order: $\vartheta^2_2-1=0$, thus one pair of solutions is
approximately $\exp(\pm \eta)$. The pure imaginary root pair
exhibits weak algebraic growth (or decay), as reflected in
$\gamma_1$.  A suitable boundary condition for (\ref{m1}) is to
match the decaying exponential and the decaying algebraic
solutions. For this purpose, we employ the following expressions
\be
& &\vartheta_1=-(A_1+2\kappa^4)^{-1/2}\nonumber
\\
& &\vartheta_2=(1+\kappa^3)(A_1+2\kappa^4)^{-1/2}-\df{A_3+B_1}{2(2\kappa^4+A_1)}-\df{b(4\kappa^4+A_1)}
{2(2\kappa^4+A_1)}\nonumber
\\
& &\gamma_1=-\df{1}{2}\kappa^2(A_1+2\kappa^4)^{-1/2};\,\,\,\,\,\,\,\,\gamma_2=-\df{3}{2}\nonumber
\en
after selecting $\ts{a}_1=\kappa^2$.
\section{Effect of the inertial critical-layer}
\subsection{Comparison with BHC}

In the case of slow waves, where $c\ll 1$, the perturbation shear stress in (\ref{m1}) can be neglected in the outer
region and thus $\Phi$ will satisfies the Rayleigh equation
\be
(U-c)(\Phi''-k^2\Phi)-U''\Phi=0\label{cl1}
\en
where now we shall assume $c=c_r$ and $c_i=0$. The corresponding expressions for unsteady waves will be given in the next subsection.

As was shown by Sajjadi (1988), the leading order solution to (\ref{cl1}) is
\be
\Phi=(U-c)e^{-k\eta}\left[{\sf A}+\Phi_cU_c'e^{k\eta_c}\int_{0}^\infty\left\{\df{1}{(U-c)^2}-1\right\}\,d\eta\right]
\label{cl2}
\en
The integral is (\ref{cl2}) is regular at the critical height and hence, by indenting the path of integration in (\ref{cl2}) under the singularity $\eta=\eta_c$, we obtain
\be
\Phi=(U-c)e^{-k\eta}\left[{\sf A}+\Phi_cU_c'e^{k\eta_c}\left(\dashint_0^\infty\left\{\df{1}{(U-c)^2}-1\right\}\,d\eta-I\right)\right]
\label{cl3}
\en
where {\sf A} is constant which can be determined from the boundary conditions and
\be
I=\lim_{\varpi\rightarrow 0}\int_{\eta_c-\varpi}^{\eta_c+\varpi}\left\{\df{1}{(U-c)^2}-1\right\}\,d\eta\label{cl4}
\en

Expanding $U(\eta)$ in a Taylor expansion in the vicinity of the critical point, setting $\eta=\eta_c\varpi e^{i\theta}$, where $\varpi\equiv c_i/U_*\ll 1$, and
\be
\tan\theta=-c_i/U_c'(\eta-\eta_c)\label{cl4a}
\en
then (\ref{cl4}) becomes
\be
I&=&\df{1}{U_c^{\prime 2}}\left\{\lim_{\varpi\rightarrow 0}\int_{\eta_c-\varpi}^{\eta_c+\varpi}\df{d\eta}{(\eta-\eta_c)^2}
+i\pi\df{U_c''}{U_c'}\right\}\nonumber\\
&=&\df{i\pi U_c''}{U_c^{\prime 3}}\label{cl5}
\en
which agrees with the result obtained by Belcher {\em et al.} (1999).

As also pointed out by Belcher {\em et al.} (1999), for a logarithmic mean velocity profile (\ref{cl4a}) yields
$\tan\theta=\varpi\eta_c/(\eta-\eta_c)$. Hence $\theta$ varies between $0$ and $\pi$ as $(\eta-\eta_c)/\ell_c$
tends to $\pm\infty$, respectively. Note that, the transition between these limiting values occurs across the
layer of thickness $\ell_c=\varpi\eta_c$. Note also, the significance of the term $iU_c''/U_c^{\prime 3}$ in the
solution for $I$ is that it leads to an out of phase contribution to the wave induced vertical velocity that gives
rise to the same contribution to the wave growth from Miles (1957) critical-layer mechanism.

The result of the present analysis confirms the earlier finding of Belcher {\em et al.} (1999) in that
Miles (1957) solution is {\em only} valid when the waves grow significantly slowly such that
\be
c_i\ll U_c'\eta_c\sim U_*\label{cl7}
\en
As in Belcher {\em et al.} (1999), our analysis also shows that when inertial effects controls the behaviour around
the critical layer, there is a smooth behaviour around the critical layer of thickness
\be
\ell_c\sim c_i/U_c'\sim\eta_cc_i/U_*\label{cl8}
\en
Hence this proves the effects of critical layer, as calculated by Miles (1957), are {\em only} valid in the
limit $c_i/U_*\downarrow 0$.
\subsection{Steady monochromatic waves}

For comparison with the waves that are unsteady,
we calculate the energy-transfer parameter due to critical layer, $\beta_c$, for steady monochromatic waves. Thus, we let $\Phi=-\mathscr{UM}$. Thus, (\ref{m1}) becomes
\be
[\nu_e({\mathscr U}{\mathscr M}''+2U'{\mathscr M}'+U''{\mathscr M})]''=ik[({\mathscr U}^2{\mathscr M}')'-
k^2{\mathscr U}^2{\mathscr M}]\label{bet1}
\en
In quasi-laminar limit the left-hand side of (\ref{bet1}) is negligible and thus (\ref{bet1}) reduces to
\be
({\mathscr U}^2{\mathscr M}')'-k^2{\mathscr U}^2{\mathscr M}=0\label{bet2}
\en
Multiplying (\ref{bet2}) by ${\mathscr M}$, integrating by parts over $0<\eta<\infty$, and invoking the
inner limits ${\mathscr M}\rightarrow a$ and ${\mathscr U}^2{\mathscr M}'\rightarrow{\mathscr P}_0$ and
a null condition at $\eta=\infty$, we obtain
\be
a{\mathscr P}_0=-\int_0^\infty{\mathscr U}^2({\mathscr M}^{\prime 2}+k^2{\mathscr M}^2)\,d\eta\label{bet3}
\en

Using the simplest admissible trial function for the variational integral (\ref{bet3}), i.e.
\be
{\mathscr M}=ae^{-k\eta/\varsigma}\label{bet3a}
\en
where $\varsigma$ is a free parameter. Substituting (\ref{bet3a}) into (\ref{bet3}) together with the approximation
${\mathscr U}\approx U_1\ln(\eta/\eta_c)$ we get
\be
\hat{{\mathscr P}}_0\equiv{\mathscr P}/kaU_1^2&=&-k(\varsigma^{-2}+1)\int_0^\infty e^{-2k\eta/\varsigma}\ln^2
(\eta/\eta_c)\,d\eta\nonumber\\
&=&-\xi_c(\varsigma^{-2}+1)\int_0^\infty e^{-st}\ln^2t\,dt\nonumber\\
&=&-\df{\varsigma+\varsigma^{-1}}{2}\left\{\df{\pi^2}{6}+\ln^2\left(\df{2\gamma\xi_c}{\varsigma}\right)\right\}\label{bet4}
\en
where $\xi_c\equiv k\eta_c$. It then follows from the variational condition $\partial\hat{{\mathscr P}}_0/\partial\varsigma=0$ that
\be
\varsigma^2=\df{L_\varsigma^2-2L_\varsigma+\pi^2/6}{L_\varsigma^2+2L_\varsigma+\pi^2/6}\label{bet5}
\en
where $L_\varsigma\equiv 2\gamma\xi_c/\varsigma$, and $\varsigma=O (1)$.

The corresponding, quasi-laminar approximation to the energy-transfer parameter may be calculated from
(\ref{cl3}), which implies $\Phi_c={\mathscr P}_c/U_c'\approx{\mathscr P}_0/U_c'$, and (\ref{cl5}), which yields
\be
\beta_c=\pi\xi_c|\Phi_c/U_1a|^2&=&\pi\xi^3_c|\hat{\mathscr P}_0|^2=\tf{1}{4}\pi(\varsigma+\varsigma^{-1})^2
\left(L_\varsigma+\tf{1}{6}\pi^2\right)^2\label{bet6}\\
&=&\pi\xi_c^3L_0^4\left[1-\left(4-\tf{1}{3}\pi^2\right)\Lambda^2+{\cal O}(\Lambda^3)\right]\label{bet7}
\en
where $\Lambda=L_0^{-1}=-\gamma-\ln(2{\sf k}), L_0=L_\delta-\ln\delta$, and 
$i\delta L_\delta/\kappa^2=(\sqrt{3}-1)+O (\delta)$. 

To obtain the corresponding expression for the component of the energy-transfer parameter, $\beta_T$, due to turbulence, we multiply (\ref{m1}) by $-{\mathscr M}$, integrating over $0<z<\infty$,
invoking the conditions
$$
{\mathscr M}=a,\hspace*{0.25cm}{\mathscr M}'=ka,\hspace*{0.25cm}{\mathscr T}'=ik[{\mathscr P}_0-kac^2]
$$
on $z=0$ and the null condition for $z\rightarrow 0$, we obtain
\begin{eqnarray}
\int_0^\infty{\mathscr M}{\mathscr T}''\,dz&=&ka[{\mathscr T}_0-i{\mathscr P}_0]+i(kac_r)^2+
\int_0^\infty{\mathscr M}''{\mathscr T}\,dz\nonumber\\
&=&i(kac_r)^2+ik\int_0^\infty
{\mathscr V}^2\left({\mathscr M}^{\prime 2}+k^2{\mathscr M}^2\right)\,dz,\nonumber
\end{eqnarray}
with $c=c_r+ic_i$. Then, in the limit as $s\equiv\rho_a/\rho_w$, where $\rho_a$ and $\rho_w$ are densities of the air and water, respectively, we obtain from
\be
\alpha+i\beta\equiv(c^2-c_w^2)/sU_1^2=
(\mathscr{P}_0+i\mathscr{T}_0)/kaU_1^2\equiv(\hat{\mathscr{P}}_0+i\hat{\mathscr{T}}_0),\label{mvd5}
\en
where $c$ is the complex wave speed, $c_w=\sqrt{g/k}$ is the speed of water waves in the absence of the airflow
above it and the suffix zero denotes evaluation at $z=0$,
\begin{eqnarray}
\alpha_T+i\beta_T=(kaU_1)^{-2}\int_0^\infty\left\{i\nu_e\left[{\mathscr V\mathscr M}^{\prime\prime 2}+
2U'{\mathscr M}/{\mathscr M}''+U''{\mathscr M}{\mathscr M}''\right]\right.& &\nonumber\\
\left.-k{\mathscr V}^2\left(
{\mathscr M}^{\prime 2}+k^2{\mathscr M}^2\right)\right\}\,dz.& &\label{mvd5a}
\end{eqnarray}
The above integral can be evaluated asymptotically\footnote{The detail evaluations may be obtained
found in the appendix of the paper by Sajjadi (2007).} whose imaginary part yields
\be
\beta_T=5\kappa^2 L_0^{-1}.\label{mvd4.10}
\en

Therefore, in summary, the energy-transfer parameter from wind to surface waves for steady monochromatic waves may be
given by the following formulae.

\begin{eqnarray}
\beta=\beta_T+\beta_c,\hspace*{1cm}
\beta_T=\df{5\kappa^2}{\Lambda},\hspace*{1cm}\beta_c=\df{5}{2}\pi\hat{W}L_0^4
\left[1-\left(4-\df{\pi^2}{3}\right)\eps^2\right]\nonumber
\end{eqnarray}
\begin{eqnarray}
\Lambda=L_0^{-1},\hspace*{0.5cm}L_0=-\gamma-\log\hat{W},\hspace*{0.5cm}
\hat{W}=kz_0e^{c_r/U_\lambda}(U_\lambda/c_r)^2,\hspace*{0.5cm}
U_\lambda=2U_*\nonumber
\end{eqnarray}
where $\kappa=0.4$ is von K\'arm\'an's constant and $\gamma=0.5772$ is Euler's constant.

\subsection{Unsteady waves}

The generalization of the results just obtained above  follows a similar development, but with the exception that $c_i\neq 0$. Here we shall present results for Stokes waves being a sum of two harmonics $(n=1, 2)$. Hence, we consider the surface wave expressed as
$$z=\Real\{ae^{ik(x-ct)}+\tf{1}{2}ka^2e^{2ki(x-ct)}\},\qquad c=c_r+ic_i$$
Note that, results for monochromatic waves follows immediately from what will be developed by setting $n=1$ and ignoring the second harmonic. 

Therefore, we begin by using the expression (5.11), but now we take
\be 
\mathscr{U}=U-c_r-ic_i\label{uw0}
\en 
Thus, the expression (5.11) now reads
\be 
a\mathscr{P}_{n0}=-\int_0^\infty[(U-c_r)^2-2ic_i(U-c_r)-c_i^2](\mathscr{M}^{\prime 2}_n
+k_n^2\mathscr{M}_n)\,d\eta\label{uw1}
\en 
where $k_1\equiv k$ and $k_2\equiv 2k$.

Once again using a similar admissible trial function as that given by (\ref{bet3a}),
namely 
$$\mathscr{M}_n=ae^{k_n\eta/\varsigma_n}\qquad(n=1, 2)$$
the variational integral (\ref{uw1}) becomes
\be 
a\mathscr{P}_{n0}&=&a^2k_n^2(\varsigma_n^{-2}+1)\left\{\int_0^\infty U_1^2\ln^2(\eta/\eta_c)e^{-2k_n\eta/\varsigma_n}\right.\,d\eta\no\\ 
&+&2ic_i\int_0^\infty U_1\ln(\eta/\eta_c)e^{-2k_n\eta/\varsigma_n}\,d\eta
+\left.c_i^2\int_0^\infty e^{-2k_n\eta/\varsigma_n}\,d\eta\right\}\no 
\en 
Evaluating the integrals, we obtain
\be 
\hat{\mathscr P}_{n0}\equiv\mathscr{P}_{n0}/k_na_nU_1^2=\df{1}{2}\left(\varsigma_n+\varsigma_n^{-1}\right)\left\{\left[\df{\pi^2}{6}+\log^2\left(\df{2k_n\gamma\eta_c}{\varsigma_n}\right)\right]-2i\hat{c}_i\log\left(\df{2k_n\gamma\eta_c}{\varsigma_n}\right)+\hat{c}_i^2\right\}\no\\\label{uw3}
\en 
where $\hat{c}_i=c_i/U_1$.

As before, applying the variational condition $\partial\hat{\mathscr P}_{n0}/\partial\varsigma_n=0$ yields
\be 
L_{n\varsigma}^2&-&2(1+i\hat{c_i})L_{n\varsigma}+\left(\hat{c}_i^2+2i\hat{c}_i+\df{\pi^2}{6}\right)\no\\
&=&\varsigma_n^2\left[L_{n\varsigma}^2+2(1+i\hat{c_i})L_{n\varsigma}+\left(\hat{c}_i^2-
2i\hat{c}_i+\df{\pi^2}{6}\right)\right]\no 
\en 
whence
\be 
\varsigma_n^2=\df{L_{n\varsigma}^2-2\frak{C}_iL_{n\varsigma}+\left(\hat{c}_i^2+2i\hat{c}_i+\df{\pi^2}{6}\right)}{L_{n\varsigma}^2+2\frak{C}_i^*L_{n\varsigma}+\left(\hat{c}_i^2-2i\hat{c}_i+\df{\pi^2}{6}\right)}\label{uw4}
\en 
where $\frak{C}_i=1+\hat{c}_i$, the superscript * denotes the complex conjugate, and
$L_{n\varsigma}\equiv 2\gamma k_n\eta_c/\varsigma_n$ where $L_{n\varsigma}=-\gamma-\log(2k_n\eta_c)$.

The expression (\ref{uw4}) may be approximated to $O (\Lambda_n^2)$,
to give 
$$\varsigma_n=\df{1-\frak{C}_i\Lambda_n}{1+\frak{C}_i^*\Lambda_n}+O (\Lambda_n^2)$$
where
$\Lambda_n\equiv L_{n0}^{-1}$, and therefore we obtain the following expression for the energy-transfer parameter to the two $(n=1, 2)$ harmonics of the wave
\be 
\beta_{nc}=\pi\left(\df{\varsigma_n L_{n\varsigma}}{2\gamma}\right)^3L_{n0}^4\left[1-\left(4-\df{\pi^2}{3}+10\hat{c}_i^2\right)\Lambda_n^2+O (\Lambda_n^3)\right]\label{uw5}
\en 
Note that, for a steady wave ($c_i=0$) and a monochromatic wave ($n=1$ and ignoring the second harmonic of the Stoke wave) the expression (\ref{uw5}) reduces to (5.16). We remark that for a steady Stokes wave we only need to assume $c_i=0$.

In a similar manner, for an unsteady wave, we adopt the complex amplitude of surface shear stress is given by (3.24) but with $\mathscr{U}$ given
by (\ref{uw0}) and the following modification for the expression for the eddy viscosity, given by (3.4), namely
\be 
\nu_{ne}=2U_*^2\{U_1[\eta^{-1}+(ik_n/\kappa^2)\ln(\eta/\eta_c)]+k_nc_i/\kappa^2\}^{-1}
\label{uw6}
\en 
Thus, upon substituting (\ref{uw6}), and using an equivalent trial function to that given by (3.25), namely
\be 
\mathscr{T}_n=\mathscr{T}_{n0}e^{-k_n\eta/\delta_n}\label{uw7}
\en  
the `unsteady' version of the integral (3.24) becomes
\be 
(\mathscr{T}_n\mathscr{T}_n')_0=-k_n\int_0^\infty[k_n/\delta_n^2+U_1^2(i\mathscr{U}^{\prime 2}-2k_n\mathscr{U}^2)]\mathscr{T}_{n0}e^{-2k_n\eta/\delta_n}\,d\eta\label{uw7a}
\en

Invoking the condition $\partial(\mathscr{T}_{n}/\mathscr{T}_{n})_0/\partial\delta_n=0$,
after substituting $\mathscr{U}$ and (\ref{uw7}), and evaluating the integral (\ref{uw7a}) we obtain
\be 
(\mathscr{T}_{n}/\mathscr{T}_{n})_0&=&\df{1}{2i\delta_n}-\df{(c_r+ic_i)^2}{4U_*^2}+\df{\delta_n}{4\kappa^2}\left(\df{1}{\delta_n}-\df{ik_n}{4\kappa^2}\right)\left(\df{\pi^2}{6}+L_{\delta_n}^2\right)\no\\
&=&\df{1}{4\kappa^2}\left[\df{2}{\delta_n}-\df{(c_r+ic_i)^2}{U_1^2}+(1+\hat{\delta_{n}})
\left(\df{\pi^2}{6}+L_{\delta_n}^2\right)\right]\label{uw8}
\en 
where $\hat{\delta_n}=i\delta_n/\kappa^2$,
\be
\left(L_{\delta_n}^2+2L_{\delta_n}+\df{\pi^2}{6}\right)\hat{\delta}_n^2+L_{\delta_n}\hat{\delta}_n-2=0\label{uw9} 
\en 
and $L_{\delta_n}=L_{n0}+\ln\delta_n$.

Solving the equation (\ref{uw9}) for $\hat{\delta}_n$ we find that
\be 
\hat{\delta_n}=(\sqrt{3}-1)(L_{n0}+\ln\delta_n)^{-1}\label{uw10}
\en 
Note that, $\delta_n$ may be complex and $|\delta_n|\ll 1$ (but strictly not equal to zero), we may use the expansion for $\ln\delta_n$, given by
$$\ln\delta_n=2\Ssf{m=1}{\infty}\df{1}{2m-1}\left(\df{\delta_n-1}{\delta_n+1}\right)^{2m-1}\doteqdot 2\df{\delta_n-1}{\delta_n+1}+O \left[\left(\df{\delta_n-1}{\delta_n+1}\right)^3\right],
\qquad 0<|\delta_n|\leq 2$$
we may cast (\ref{uw10}) as
\be 
\hat{\delta}_n=(\sqrt{3}-1)\left[L_{n0}+2\left(\df{\delta_n-1}{\delta_n+1}\right)\right]^{-1}\label{uw11} 
\en 

Hence, the asymptotic evaluation of integrals in (\ref{uw7a}) yields to the following
expression (for further details see the appendix of Sajjadi (2007))
\be 
\beta_n=2\kappa^2\left[1+(\sqrt{3}+1)\left(1-\mathscr{E}_{\delta_n}\right)L_{|\delta_n|}-(\sqrt{3}-1)\left(L_{|\delta_n|}+4\ln 2\right)\right. & &\no\\
-\left. 4\left(\mathscr{E}_{\delta_n}-\mathscr{E}_{2\delta_n}\right)\right]-4\delta_{ni}L_{n0}^2 & &\label{uw12}
\en 
where $\delta_{ni}=\mathscr{I}\{\delta_n\}$,
\be
\mathscr{E}_p=\mathscr{E}\left(\df{2a_1}{ipL_p}\right)\approx 1-\df{ipL_p}{2a_1},\qquad
L_p=-\gamma-\ln\left(2\sf{k}{p}\right)=L_0+\ln p\no 
\en 
and 
$$\mathscr{E}(\Theta_n)=\Theta_ne^{\Theta_n}\int_0^\infty t^{-1}e^{-t}\,dt=\Theta_ne^{\Theta_n}{\sf E}_1(\Theta_n)$$
From these expressions we obtain (see Sajjadi (2007)) we obtain
$$\beta_{nc}=\df{1}{2}(\Lambda_n+\Lambda_n^{-1})\left\{L_{\varsigma_n}^2+\df{\pi^2}{6}\right\},\qquad\beta_{nT}=\df{4\kappa^2}{\Lambda_n}$$
which are in reasonable agreement with (5.19) and (5.24).

\section{Results and conclusions}

In figure 5, we show comparison of the energy-transfer rate, $\beta$, between the present result for a monochromatic unsteady (growing) wave, both analytically and numerically, and those calculated by Miles (1957) and Janssen (1991) for the steady wave counterpart.  Miles and Janssen both assume that the drag $C_D$, and thence $\beta$, is dominated by the limiting inviscid wave growth mechanism, thus their formulation is independent of $c_i$.  In contrast, the present calculation is for a viscous unsteady (growing) wave, where $c_i/U_*=0.01$ and $kz_0=10^{-4}$.

We emphasize that the various models, such as those by Belcher \& Hunt (1993), Mastenbroek {\em et al.} (1996), Cohen (1997) etc., all generally agree with our numerical simulations performed using Launder, Reece \& Rodi (1975) Reynolds-stress closure scheme for the energy  transfer parameter, $\beta$, shown in figure 5. This shows consistency between these  models and the unimportance of very small $c_i$ for which viscous processes are significant.

Figure 6 shows comparison of $\beta_c$ as a function of wave age $c_r/U_1$, calculated according to (Conte \& Miles 1959) for numerical solution of inviscid Orr-Sommerfeld equation, against the numerical solution of equation (1.6) for $c_i/U_*=0.01$, $kz_0=10^{-4}$ and $\nu_e\neq 0$. We remark that increasing $c_i/U_*$ from 0.01 to 0.1 (not shown here) makes no significant difference in the magnitude of $\beta_c$.  We conclude therefore for a finite value of $\nu_e$ the right-hand side of equation (1.6) is dominant and therefore the magnitude of $\beta_c$, calculated from the solution of (1.6), is practically zero over a wide range of the wave age, in particular for a `young' wave, where $c_r/U_1<2.$ We thus conclude that the critical-layer mechanism plays an insignificant role for $c_r/U_1<9$, and very little effect for $9\leq c_r/U_1\leq 10.5$. 

\begin{figure}
   \begin{center}
\includegraphics[width=12cm]{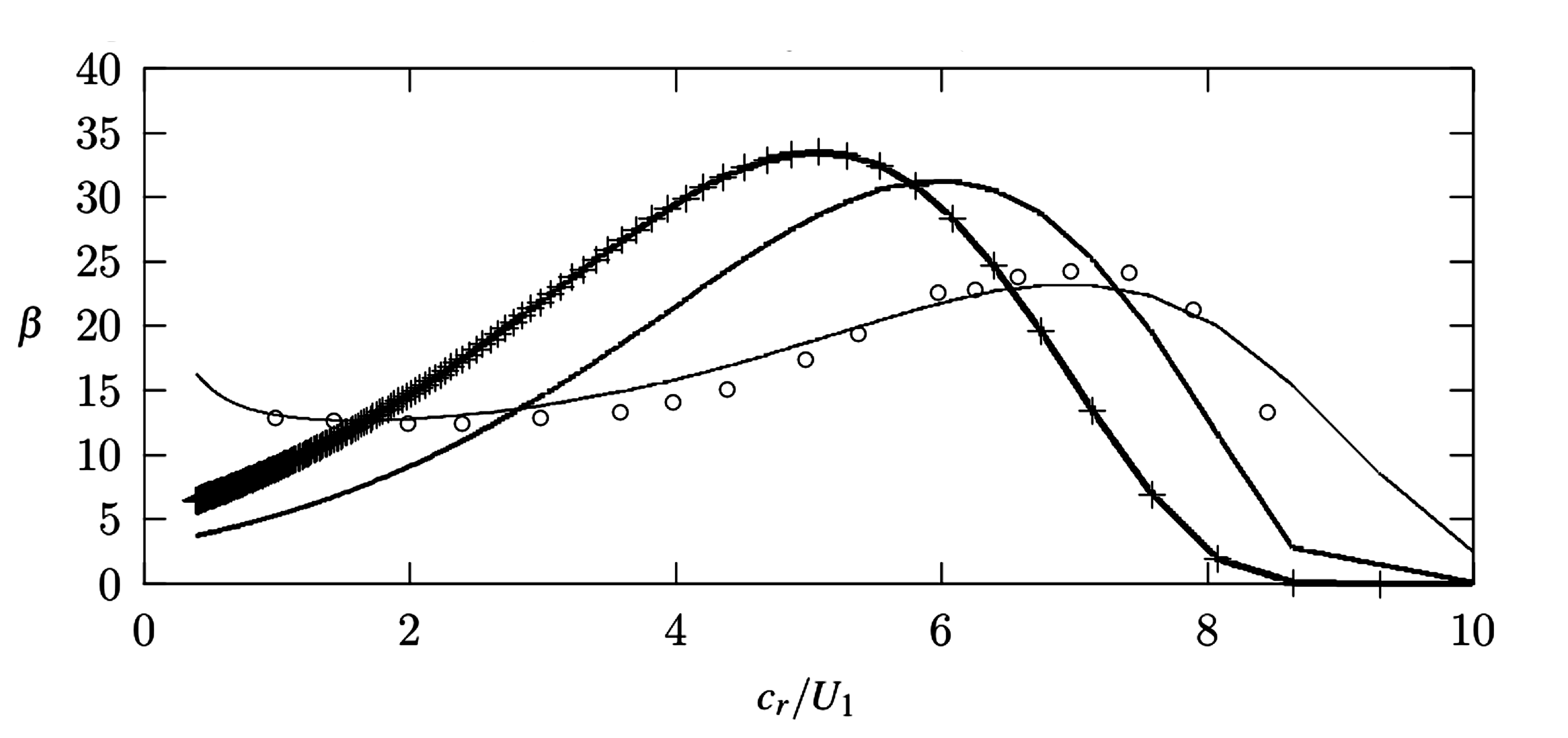}
\caption{\footnotesize Total energy transfer parameter, $\beta$, using different models for critical layer and sheltering mechanisms
for unsteady waves (where $c_i\ll U_*$) as a function of the wave age $c_r/U_1$.
 +++++, Miles (1957) calculation ($c_i=0, \nu_e=0$) from his formula: 
$\beta=\pi\eta_c\left\{\tf{1}{6}\pi^2+\log^2(\gamma\eta_c)+2\sum_{n=1}^{\infty}\tf{(-1)^n\eta_c^n}
{n!n^2}\right\}^2$, where $\eta_c=kz_c$ is the critical height $\eta_c=\Omega(U_1/c_r)^2e^{c_r/U_1}$ and $\Omega=gz_0/U_1^2$ is the Charnock's (1955) constant. Thick solid line, parameterization of Miles (1957) formula, for $c_i=0, \nu_e=0$, given by Janssen (1991):
$\beta=1.2\kappa^{-2}\eta_c\log^4\eta_c$, where
$\eta_c=\min\left\{1,kz_0e^{[\kappa/(U_*/c+0.011)]}\right\}$.
Thin solid line, present formulation: ($\beta_T+\beta_c$) for $c_i\neq 0, \nu_e\neq 0$. $\circ$, Numerical simulation using Launder, Reece \& Rodi (1975) Reynolds-stress closure model for $c_i\neq 0, \nu_e\neq 0$. Note that, $\beta$ for Miles and Janssen is equivalent to $\beta_c$ in our notation. Taken from Sajjadi {\em et al.} (2014).}
\end{center}
\end{figure}

We remark that these parameterizations have been incorporated and tested in a spectral
wave models, WaveWatch and WindWave, which shows superior results when compared with field data (see Fitzpatrick {\it et al.} 2002 and Sajjadi {\it et al.} 2002).

\begin{figure}
   \begin{center}
\includegraphics[width=12cm]{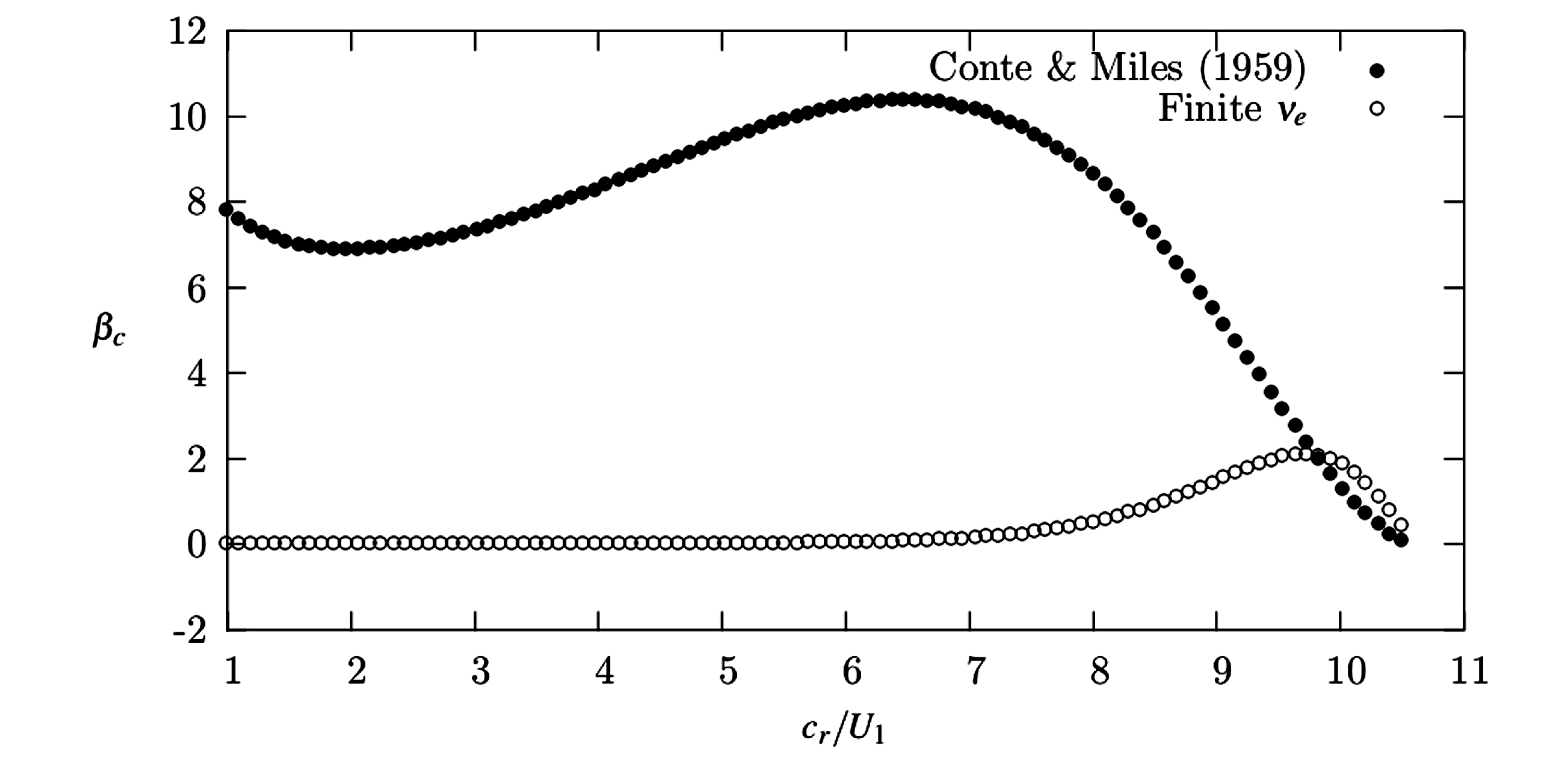}
\caption{\footnotesize Component of energy transfer parameter, $\beta_c$, by different models of  critical layer mechanisms
for unsteady waves (where $c_i\ll U_*$) as a function of the wave age $c_r/U_1$. $\bullet$, numerical solution of inviscid Orr-Sommerfeld equation by Conte \& Miles (1959) for $c_i=0$ and $\nu_e=0$ using the singular critical layer approach; $\circ$ numerical solution of equation (1) for $c_i\neq 0$ and $\nu_e\neq 0$. Taken from Sajjadi {\em et al.} (2014).}
\end{center}
\end{figure}

In conclusion we adopted an asymptotic multi-deck solution for turbulent shear flows over unsteady surface waves, in the limits of low 
turbulent stresses and small wave amplitude.  The structure of the flow is defined, using an eddy-viscosity turbulence model, in 
terms of asymptotically-matched thin-layers, namely the surface layer and a critical layer. Solutions for both inner and outer
regions are constructed through an interpolation between an inner, mixing-length and an outer, rapid-distortion approximations. 
The results particularly demonstrate the physical importance of the singular flow features and physical implications of the elevated
critical layer in the limit of the unsteadiness tending to zero.  These agree with the variational mathematical solution of Miles (1957) 
for small but finite growth rate. However, the results obtained here, are not consistent physically or mathematically with his analysis in the limit of
growth rate tending to zero.  In the present study it is shown that in the limit of zero growth rate the effect of the elevated critical 
layer is eliminated by finite turbulent diffusivity, so that the perturbed flow and the drag force on determined by the asymmetric and
sheltering flow in the surface shear layer and its matched interaction with the upper region, as physically demonstrated by Sajjadi, Hunt and Drullion (2014). The results for an unsteady monchromatic waves is also extended to those growing Stokes waves. Thus, estimation can be made as to what precentage of total energy transfer from wind goes to each harmonic of a Stokes wave.  

\section*{References}
\begin{description}
\item{[1]} BELCHER, S.E. AND HUNT, J.C.R. (1993) Turbulent shear flow over slowly moving waves, J. Fluid Mech., 251, 109.

\item{[2]} BELCHER, S.E., HUNT, J.C.R. AND COHEN, J.E. (1999) Turbulent flow over growing waves. In Proceedings of IMA Conference on Wind over Waves, (Eds. S.G. Sajjadi, N.H. Thomas AND J.C.R. Hunt), 19–30, Oxford University Press.

\item{[3]} CHARNOCK, H. (1955) Wind stress on a water surface, Q. J. R. Meteorol. Soc., 81, 639.

\item{[4]} CONTE, S.D. AND MILES, J.W. (1959) On the numerical integration of the Orr-Sommerfeld equation, J. Soc. Indust. Appl. Math., 7, 361.

\item{[5]} FITZPATRICK, P., MOSTOVOI, G., LI , Y., BETTENCOURT, M. AND SAJJADI , S.G. (2002) Coupling of COAMPS and WAVEWATCH with Improved Wave Physics, DoD High Performance Computing Modernization Program
Programming Environment and Training Report No. 0005.

\item{[6]} IERLEY, G. AND MILES, J.W. (2001) On Townsend's rapid-distortion model of the turbulent-wind-wave problem, J. Fluid. Mech., 435. 175. 

\item{[7]} JANSSEN, P.A.E.M. (1991) Quasi-linear theory of wind-wave generation applied to wave forecasting, J. Phys. Oceanog., 21, 1631.

\item{[8]} JEFFREYS, H. (1925) On the formation of water waves by wind, Proc. R. Soc. London Ser. A, 107, 189.

\item{[9]} LAUNDER, B.E., REECE, G.J. AND RODI, W. (1975) Progress in the development of a Reynolds-stress turbulence closure, J. Fluid Mech., 68, 537.

\item{[10]} Lighthill, M.J. (1957) The fundamental solution for small steady three-dimensional parallel shear flow. J. Fluid. Mech. 14, 385.

\item{[11]} OLVER , F.W.J. (1974) Asymptotics and special functions, Academic Press.

\item{[12]} MASTENBROEK, C., MAKIN, V.K., GARAT, M.H. AND G IOVANANGELI, J.P. (1996) Experimental evidence of the rapid distortion of turbulence in air flow over waves, J. Fluid Mech., 318, 273.

\item{[13]} MILES, J.W. (1957) On the generation of surface waves by shear flows, J. Fluid Mech., 3, 185.

\item{[14]} MILES, J.W. (1996) Surface-wave generation: a viscoelastic model, J. Fluid Mech., 322, 131.

\item{[15]} ROTTA, J. (1950) Das in Wandn\"a g\"ultige Geschwindigkeitsgesetz turbulenter Str\"omungen. Archive of Applied Mechanics, 18, 277.

\item{[16]} SAJJADI, S.G. (1988) Shearing flows over Stokes waves. Department of Mathematics Internal Report, Coventry Polytechnic, UK.

\item{[17]} SAJJADI, S.G. (1998) On the growth of a fully nonlinear Stokes wave by turbulent shear flow. Part 2: Rapid distortion theory, Math. Eng. Ind., 6, 247.

\item{[18]} SAJJADI, S.G., BETTENCOURT, M.T., F ITZPATRICK, P.J., MOSTOVOI, G. AND LI, Y. (2002) Sensitivity of a coupled tropical cyclone/ocean wave simulation to different energy transfer schemes. In Proceeding of 25th Conference on Hurricanes and Tropical Meteorology. 15D.5.

\item{[19]} SAJJADI, S.G. (2007a) Interaction of turbulence due to tropical cyclones with surface waves. Adv. Appl. Fluid Mech. 1, 101.

\item{[20]} SAJJADI, S.G. (2010) On asymptotic solution of $\mathscr{T} − (ik\mathscr{U}/\nu k)\mathscr{T}=0$ for large $k$, Adv. Appl. Fluid Mech., 7, 31.

\item{[21]}  SAJJADI, S.G, HUNT, J.C.R. AND DRULLION, F. (2014) Asymptotic multi-layer analysis of wind over unsteady monochromatic surface waves. J. Eng. Math. 84, 73.

\item{[22]} STEWARTSON, K. (1974) Boundary layers on flat plates and related bodies. Adv. Appl. Mech. 14, 145.

\item{[23]} TOWNSEND, A.A. (1972) Flow in a deep turbulent boundary layer over a surface distorted by water waves, J. Fluid Mech., 55, 719.

\end{description}
\end{document}